\documentclass[11pt]{article}
\usepackage{gbmacros}
\usepackage{cite}
\usepackage{times,mathpazo,charter,graphicx,tocloft}
\usepackage[bookmarks=false,colorlinks=true,citecolor=red,linkcolor=blue,pdfstartview={FitH}]{hyperref}
\usepackage{verbatim}
\usepackage{subcaption}
\usepackage{pdfpages}
\usepackage{amsmath, amssymb, graphics, setspace}
\usepackage{fullpage}

\advance\textwidth 8mm
\advance\hoffset -4mm
\advance\textheight 13mm
\advance\voffset -6mm

\def\maketitle{\par\noindent{\LARGE\bf\sffamily\thetitle}\\[1.6ex]
{\large\theauthor}\\[0.6ex]
\textit{\thetextinfo}\\[0.2ex]
{\small\today}\par\vglue1.4\bigskipamount}
\def\title#1{\def\thetitle{#1}}
\def\author#1{\def\theauthor{#1}}
\def\textinfo#1{\def\thetextinfo{#1}}
\allowdisplaybreaks

\makeatletter
\def\d{\mathrm{d}}
\def\cn{\mathop{\rm cn}\nolimits}
\def\sn{\mathop{\rm sn}\nolimits}
\def\dn{\mathop{\rm dn}\nolimits}
\def\K{\mathsf{K}}
\def\E{\mathsf{E}}
\let\~=\tilde
\def\DDy#1{\frac{D#1}{Dy}}
\def\ddy#1{\partialderiv{#1\!}{y}\,}
\makeatother

\def\be{\begin{equation}}
\def\ee{\end{equation}}
\def\bse{\begin{subequations}}
\def\ese{\end{subequations}}

\advance\parskip 2pt

\usepackage{color}

\begin{document}

\title{Whitham modulation theory for the\\[0.4ex]Kadomtsev-Petviashvili equation} 
\author{Mark J. Ablowitz$^1$, Gino Biondini$^{2,3}$ and Qiao Wang$^2$}
\textinfo{%
$^1$ Department of Applied Mathematics, University of Colorado, Boulder, CO 80303\\
$^2$ Department of Mathematics, State University of New York, Buffalo, NY 14260\\
$^3$ Department of Physics, State University of New York, Buffalo, NY 14260
}
\date{ }
\maketitle

\noindent\begin{minipage}{0.905\textwidth}
\small
\textbf{Abstract.}
The genus-1 KP-Whitham system is derived for both variants  of the Kadomtsev-Petviashvili (KP) equation;
namely the KPI and KPII equations.
The basic properties of the KP-Whitham system, including symmetries, exact reductions, and its possible complete integrability, 
together with the appropriate generalization of the one-dimensional Riemann problem for the Korteweg-deVries equation are discussed.
Finally, the KP-Whitham system is used to study the linear stability properties of the genus-1 solutions of the 
KPI and KPII equations; it is shown that all genus-1 solutions of KPI are linearly unstable while all genus-1 solutions of KPII are linearly stable within the context of Whitham theory.


\end{minipage}

\vskip2\bigskipamount


\section{Introduction}
\label{s:intro}

Small-dispersion limits and dispersive shock waves (DSWs) continue to be the subject of considerable research; 
see for example~\cite{physd2016preface,deng2016physd,elhoefer2016physd,elgrimshaw2002,grimshawyuan,kamchatnov2016pkdv,marchantsmyth,D6,D7,D8,biondinitrilloprl2016,D9} and references therein.
The prototypical example where DSWs arise is the Korteweg-de Vries (KdV) equation
\begin{align}
\label{e:kdv}
u_t + 6 u u_x + \epsilon^2 u_{xxx} = 0\,,
\end{align}
with a unit step initial condition (IC), namely $u(x,0) = 1$ for $x<0$ and $u(x,0)=0$ for $x\ge 0$.
In 1974, using the averaging method pioneered by Whitham~\cite{Whitham}, Gurevich and Pitaevskii gave a detailed description of the associated DSW\cite{GP1974}. 
Over the last forty years, there have been numerous studies regarding small dispersion limits and DSWs.
Most analytical results are limited to (1+1)-dimensional  partial differential equations (PDEs), however.
This work is concerned with the study of DSWs associated with a (2+1)-dimensional PDE, 
namely, the celebrated Kadomtsev--Petviashvili (KP) equation
\begin{align}
\label{e:kp}
(u_t + 6 u u_x + \epsilon^2 u_{xxx})_x + \lambda u_{yy} = 0\,,
\end{align}
where $\epsilon > 0$ is a small parameter. 
The case $\lambda = -1$ is known as the KPI equation, whereas the case $\lambda = 1$ is known as the KPII equation. 
Equation~\eqref{e:kp}, which was first derived by Kadomtsev and Petviashvili~\cite{KP} in the context of plasma physics,
is a universal model for the evolution of weakly nonlinear two-dimensional long water waves of small amplitude, 
and arises in a variety of physical settings.
In the context of water waves, KPI describes the case with weak surface tension and KPII describes the case with strong surface tension cf. \cite{AS1981}.
The KP equation is also the prototypical (2+1)-dimensional integrable system.  
As such, it has been heavily studied analytically over the last forty years;
see for example \cite{AS1981,AblowitzClarkson,BBEIM,Bogaevskii,gbsc2006jmp,gbyk2003jpa,ep4,hakkaev,haragus,infeld,
ep5,freemannimmo,klein,krichever,ep7,ep8,b3,kuznetsov,ep9,ep10,b4,b5,gb2007prl,ep11,ep12,ep13} and references therein.

The behavior of solutions of the KPI and KPII equations with small dispersion was recently studied numerically by Klein et al~\cite{klein};
see also~\cite{grava} for a study of shock formation in the dispersionless KP.
Even though there have been a few works about the derivation of a Whitham system for the KP equation,
\cite{Bogaevskii,infeld,krichever}, 
to the best of our knowledge there are no studies in which such systems were written in Riemann-type variables,
nor studies regarding the use of these systems to study DSWs. 

In this work 
we begin a program of study aimed at overcoming 
these deficiencies 
by generalizing Whitham modulation theory to (2+1)-dimensional PDEs
to study (2+1)-dimensional DSWs.
A first step towards this goal was recently presented in~\cite{ckdv}, where the cylindrical reduction of the KP equation~\eqref{e:kp} was studied. 
More specifically, the authors of~\cite{ckdv} considered the special case in which the solution of the KP equation depends on $x$ and $y$ only through the similarity variable 
$\eta = x + P(y,t)$. 
In particular, with the special choice of a parabolic initial front, 
namely, $P(y,0) = c y^2/2$, \eqref{e:kp} reduces to the cylindrical KdV (cKdV) equation
\begin{align}
\label{e:ckdv}
u_t + 6 u u_\eta + \frac{\lambda c}{1+2 \lambda c t} u+ \epsilon^2 u_{\eta\eta\eta} =0\,,
\end{align}
the DSW behavior of which was then studied in \cite{ckdv} with step-like initial data.

In this work we generalize the above results to the fully (2+1)-dimensional case.
More precisely, we derive the (2+1)-dimensional Whitham system of the KP equation using the method of multiple scales
(e.g., as in Luke~\cite{Luke}). 
The main result of this work is the $5\times5$ system of (2+1)-dimensional hydrodynamic-type equations (see also item 10 on section~\ref{s:remarks})
\bse
\label{e:kpwhitham}
\begin{align}
&\partialderiv{r_j}{t} + (V_j + \lambda q^2)\,\partialderiv{r_j}{x} + 2 \lambda q  \DDy{r_j} + \lambda \nu_j \DDy{q}
  { + \lambda \DDy{p}}
  = 0\,, \qquad j=1,2,3,\\
&\partialderiv{q}{t} + (V_2 + \lambda q^2)\,\partialderiv{q}{x} + 2 \lambda q \DDy{q} 
+ \nu_{4.1} \DDy{r_1} + \nu_{4.3} \DDy{r_3} = 0\,,  \\
\label{e:pcontr}
&{
\partialderiv{p}{x} - (1-\alpha)\DDy{r_1} - \alpha\DDy{r_3} + \nu_5 \partialderiv{q}{x} = 0\,,
}
\end{align}
\ese
where all the coefficients are given explicitly~{by equation~\eqref{e:kpwhithamcoeffs}} in Section~\ref{s:derivation},
and where for brevity we used the ``convective'' derivative
\begin{equation}
\label{e:DDy}
\frac{D}{D y} = \partialderiv{ }y - q \partialderiv{ }x\,,
\end{equation}
which will be used throughout this work.
The system~\eqref{e:kpwhitham} describes the slow modulations of the periodic solutions of the KP equation~\eqref{e:kp}, 
and is the (2+1)-dimensional generalization of the 
Whitham systems for the KdV equation~\eqref{e:kdv} and cKdV equation~\eqref{e:ckdv}.
Hereafter we refer to \eqref{e:kpwhitham} as the \textit{KP-Whitham system}.

The outline of this work is the following. 
In section~\ref{s:derivation} we derive the KP-Whitham system~\eqref{e:kpwhitham} of modulation equations using a multiple scales approach,
{and we discuss how our results compare to previous studies in the literature.}
In section~\ref{s:properties} we discuss basic properties of the KP-Whitham system~\eqref{e:kpwhitham} such as symmetries and exact reductions,
{as well as the formulation of well-posed initial value problems for it, 
including the (2+1)-dimensional generalization of the Riemann problem for the KdV equation.}
In section~\ref{s:stability} we use the KP-Whitham system~\eqref{e:kpwhitham} to study the linear stability properties of genus-1 solutions 
and the DSW of the KdV equation within the KPI and KPII equations. 
{We also compare the analytical predictions to direct numerical calculations 
of the spectrum of the linearized KP equation around a periodic solution, showing excellent agreement.}
In the soliton limit, which coincides with the soliton front of the DSW, the growth rate {from Whitham theory also} 
agrees with {analytical result obtained from a direct linearized stability analysis of the soliton with respect to} 
transverse perturbations.
In the Appendix we give a brief review of the Whitham system for the KdV equation, 
a discussion of direct stability analysis for the genus-1 solution of the KP equation,
further details on the regularization of the KP-Whitham system
and the explicit values of various quantities appearing in the following sections.

\section{Derivation of the Whitham system for the KP equation}
\label{s:derivation}

\subsection{The multiple scales expansion}
\label{s:multiscale}

The KP equation~\eqref{e:kp} was originally derived in the form~\cite{KP}
\bse
\label{e:evol}
\begin{align}
&u_t + 6 u u_x + \epsilon^2 u_{xxx} + \lambda v_y = 0\,, 
\label{e:kp1}
\\
&v_x = u_y\,.
\label{e:kp2}
\end{align}
\ese
Here we use the method of multiple scales to derive modulation equations for the traveling wave (i.e., elliptic, or genus-1) solutions of the KP equation in the above form via Whitham modulation theory. 
The result will be {the five} (2+1)-dimensional quasi-linear first order PDEs~\eqref{e:kpwhitham}
that describe the evolution of the parameters of the traveling wave solution of the KP equation.

To apply the method of multiple scales, we start by looking for the solution of KP equation in the form of 
$u = u(\theta, x, y, t)$ with the rapidly varying variable $\theta$ defined from
\begin{align}
\label{e:thetaderiv}
\theta_x = {k(x,y,t)}/{\epsilon}\,, \qquad 
\theta_y = {l(x,y,t)}/{\epsilon}\,, \qquad 
\theta_t = -{\omega(x,y,t)}/{\epsilon}\,,
\end{align}
where $k,~l$ and $\omega$ are the wave numbers and frequency, respectively, which are assumed to be slowly varying functions of $x$, $y$ and $t$.
Imposing the equality of the mixed second derivatives of $\theta$ then leads to the compatibility conditions
\vspace*{-0.4ex}
\bse
\label{e:compatibility}
\begin{align}
k_t + \omega_x = 0\,, \label{e:eq1} \\
l_t + \omega_y = 0\,, \label{e:eq2} \\
k_y - l_x = 0\,. \label{e:constraint}
\end{align}
\ese
Equations~\eqref{e:eq1} and~\eqref{e:eq2} are usually referred as the equations of conservation of waves.
They provide the first and the second modulation equations. 
Note also that~\eqref{e:eq1} and~\eqref{e:eq2} automatically imply that, if~\eqref{e:constraint} is satisfied at $t=0$, it is satisfied for all $t>0$.
This fact will be used later to simplify the Whitham system. 

With these fast and slow variables, the system~\eqref{e:evol} transforms according to 
\begin{align}
\label{e:deriv}
\partialderiv{ }x \mapsto \frac{k}{\epsilon} \frac{\partial}{\partial\theta} + \partialderiv{ }x\,, 
\quad \partialderiv{ }y \mapsto \frac{l}{\epsilon} \frac{\partial}{\partial\theta} + \partialderiv{ }y\,,
\quad \partialderiv{ }t \mapsto -\frac{\omega}{\epsilon} \frac{\partial}{\partial\theta} + \partialderiv{ }t\,,
\end{align}
which yields
\bse
\label{e:kpasymp}
\begin{align}
&\frac{1}{\epsilon}\bigg(-\omega \frac{\partial u}{\partial \theta} + 6 k u \frac{\partial u}{\partial \theta} + k^3 \frac{\partial^3 u}{\partial \theta^3} + \lambda l \frac{\partial v}{\partial \theta} \bigg) + \bigg(  \partialderiv{u}{t} + u \partialderiv{u}{x} + 3 k k_x \frac{\partial^2 u}{\partial \theta^2} + 3 k^2 \frac{\partial^3 u}{\partial \theta^2 \partial x} + \lambda \partialderiv{v}{y} \bigg)
\nonumber\\
&\kern4em{ } +\epsilon \bigg(k_{xx} \frac{\partial u}{\partial \theta} + k_x \frac{\partial^2 u}{\partial \theta \partial x} + k \frac{\partial^3 u}{\partial \theta \partial^2 x} + 2 k \frac{\partial^3 u}{\partial^2 \theta \partial x} \bigg) + \epsilon^2 \frac{\partial^3 u}{\partial^3 x} = 0\,, 
\\
&\frac1\epsilon \bigg( k\frac{\partial v}{\partial\theta} - l \partialderiv u\theta \bigg) + \bigg( \partialderiv vx - \partialderiv uy \bigg ) = 0\,.
\end{align}
\ese
We then look for an asymptotic expansion of $u$ and $v$ in powers of $\epsilon$ as
\bse
\label{e:uvasymp}
\begin{align}
u = u^{(0)}(\theta, x,y,t) + \epsilon u^{(1)}(\theta, x,y,t) + O(\epsilon^2)\,, \\
v = v^{(0)}(\theta, x,y,t) + \epsilon v^{(1)}(\theta, x,y,t) + O(\epsilon^2)\,.
\end{align}
\ese
Grouping the terms in like powers of $\epsilon$ yields leading-order and higher-order problems.
It is sufficient to only consider the first two orders.

The leading terms, found at $O(1/\epsilon)$, yield
\bse
\label{e:leading}
\begin{align}
\label{e:leading1}
&-\omega u^{(0)}_\theta + 6 k u^{(0)} u^{(0)}_\theta + k^3 u^{(0)}_{\theta\theta\theta} + \lambda l v^{(0)}_\theta = 0\,, \\
\label{e:leading2}
&k v^{(0)}_\theta = l u^{(0)}_\theta\,.
\end{align}
\ese
Equations~\eqref{e:leading} can be written in compact form as
\begin{align}
\label{e:leading3}
\@M_0~\@u^{(0)} = \@0
\end{align}
where $\@u^{(j)} = (u^{(j)},v^{(j)})^T$, 
$\@0$ is the zero vector and 
$\@M_0 = \@M\,\partial_\theta$,
with
\begin{gather}
\@M = \begin{pmatrix} 
  \mathcal{L}     & \lambda qk\\ 
  \lambda qk & -\lambda k
\end{pmatrix}\,,
\qquad
\mathcal{L} = -\omega + 6 k u^{(0)} + k^3 \partial^2_\theta \,,
\end{gather}
and where 
we introduced the dependent variable
\begin{align}
q(x,y,t) = l/k\,, 
\label{e:qdef}
\end{align}
which will be used throughout the rest of this work.
Integrating~\eqref{e:leading2} we obtain
\begin{align}
\label{e:v0def}
v^{(0)} = q u^{(0)} + p\,,
\end{align} 
where 
$p(x,y,t)$ is an integration constant that is up to this point arbitrary, and must be determined at higher order in the expansion.

Next we look at the $O(1)$ terms, which yield
\bse
\label{e:next}
\begin{align}
&u^{(0)}_t - \omega u^{(1)}_\theta + 6 u^{(0)} u^{(0)}_x + 6 k (u^{(0)} u^{(1)})_\theta + 3 k k_x u^{(0)}_{\theta\theta} + 3 k^2 u^{(0)}_{\theta\theta x} + k^3 u^{(1)}_{\theta\theta\theta} + \lambda (v^{(0)}_y + l v^{(1)}_\theta) = 0\,, \\
&v^{(0)}_x + k v^{(1)}_\theta = u^{(0)}_y + l u^{(1)}_\theta\,.
\end{align}
\ese
Again we can write the above equations in vector form as 
\begin{align}
\label{e:next2}
\@M_1~\@u^{(1)} = \textbf{G}
\end{align}
where
$\@G = (g_1,g_2)^T$,  
$\@M_1 = \partial_\theta \@M$, 
and
\bse
\label{e:Gdef}
\begin{align}
  g_1 &=
    - u^{(0)}_t - 6 u^{(0)} u^{(0)}_x - 3 k k_x u^{(0)}_{\theta\theta} - 3 k^2 u^{(0)}_{\theta\theta x} - \lambda v^{(0)}_y \,,\\
  g_2 & = 
    \lambda v^{(0)}_x - \lambda u^{(0)}_y \,. 
\end{align}
\ese
Note that the matrix differential operator $\@M_1$ is a total derivative in $\theta$.
We will see in the following section that the solution $\@u^{(0)}$ of~\eqref{e:leading}, is periodic, namely,
\begin{align*}
\mathbf{u}(\theta + P) = \mathbf{u}(\theta)\,,
\end{align*}
%
where the period $P$ is yet to be determined.
Integrating~\eqref{e:next2} and imposing the absence of secular terms, we then obtain the {vector} condition
\begin{align}
\int_{0}^{P} \textbf{G} \,d\theta = 0\,,
\label{e:periodicity}
\end{align}
{which will provide two more modulation equations.}	
To obtain the last modulation equation, note that
the Fredholm solvability condition for the inhomogeneous problem~\eqref{e:next2} is
\begin{align*}
\int_{0}^{P} {\@w \cdot \@G}~ d\theta = 0\,,
\end{align*}
where $\@w$ is any solution of the homogeneous problem for the adjoint operator at  $O(1)$. 
That is, 
\begin{align*}
\@M_1^\dagger \mathbf{w} = \@0\,,
\end{align*}
where $\dagger$ denotes the Hermitian conjugate.
Using periodicity, it is easy to verify that 
$\mathcal{L}$ is self-adjoint, which implies $\@M_1^\dagger = - \@M_0$.
Therefore $\@w = \@u^{(0)}$, and the solvability condition becomes
\begin{align}
\label{e:fredholm}
\int_{0}^{P} {\mathbf{u}^{(0)} \cdot \mathbf{G}} ~d\theta = 0\,,
\end{align}
which yields the last modulation equation.
Summarizing, we have {five} modulation equations: \eqref{e:eq1}, \eqref{e:eq2} and \eqref{e:fredholm}
and the two-component periodicity condition~\eqref{e:periodicity}.

\subsection{Modulation equations for the parameters of the elliptic solutions}
\label{s:elliptic}

Here we obtain the explicit form for the {five} modulation equations,
which will provide PDEs for the evolution of the characteristic parameters of the traveling wave solution of the KP equation.

We begin by going back to the $O(1/\epsilon)$ term~\eqref{e:leading}. 
{
Using~\eqref{e:leading2} to eliminate $v^{(0)}_\theta$, we can rewrite~\eqref{e:leading1} as
}
\begin{align}
\label{e:ellipticeqn}
k^2 u_{\theta\theta\theta}^{(0)} + 6 u^{(0)} u_\theta^{(0)} -V u_\theta^{(0)}  = 0\,,
\end{align}
where 
\begin{align}
\label{e:omegadef}
V = \frac{\omega}{k} - \lambda q^2\,.
\end{align}
The solution of~\eqref{e:ellipticeqn} is (e.g., see \cite{AblowitzClarkson})
\begin{align}
\label{e:ellipticsoln}
u^{(0)}(\theta,x,y,t) = a(x,y,t) + b(x,y,t)\cn^2(\Xi, m)\,,
\end{align}
where $\cn(\cdot,m)$ is one of the Jacobi elliptic functions \cite{nist}, $m$ is the elliptic parameter (i.e., the square of the elliptic modulus),
\begin{align}
\label{e:abxidef}
\Xi(\theta) = 2\K(\theta - \theta_0)\,, \qquad
a = \frac{V}{6} - \frac{2 b}{3} + \frac{b}{3 m}\,, \qquad b = 8 m k^2 \K^2\,,  
\end{align}
and $\K = K(m)$ {and $\E = E(m)$ are} the complete elliptic integrals of the first {and second} kind, {respectively}~\cite{nist}. 
The solution~\eqref{e:ellipticsoln} can be verified by direct substitution by noting that
\begin{align*}
&u_\theta^{(0)} = -4b\K \cn(\Xi, m) \sn(\Xi, m) \dn(\Xi, m)\,, \\
&u_{\theta\theta\theta}^{(0)} = 64b\K^3 \cn(\Xi, m) \sn(\Xi, m) \dn(\Xi, m) (1 - 2 m - 3 m \cn^2(\Xi, m))\,.
\end{align*}
When $a$, $b$, $m$, $k$, $q$ and $\omega$ are independent of $x$, $y$ and $t$, \eqref{e:ellipticsoln} is the well-known exact cnoidal wave solution of the KP equation. 
(Note that, even though six constants appear in $u^{(0)}$, there are only four independent parameters.) 
Conversely, if these quantities are slowly varying functions of $x$, $y$ and $t$, one obtains a slowly modulated elliptic wave.
In this case, the four independent parameters satisfy a system of nonlinear PDEs of hydrodynamic type.
More precisely, the solution~\eqref{e:ellipticsoln} is determined (up to a constant $\theta_0$) by the four independent parameters $V,~b/m,~m$ and~$q$, 
and we next show that the evolution of these parameters is 
uniquely determined by the modulation equations derived above.

As the Jacobi elliptic function $\cn(u,m)$ has period $2\K$, the elliptic solution $u^{(0)}$ has period $1$ as a function of $\theta$, 
i.e., $P=1$ in~\eqref{e:periodicity} and~\eqref{e:fredholm}.
Recall that the five modulation equations are given by 
\eqref{e:eq1}, \eqref{e:eq2} and \eqref{e:fredholm}
and the two-component condition~\eqref{e:periodicity}.
Using~\eqref{e:v0def} to eliminate $v^{(0)}$ and substituting~\eqref{e:Gdef} into~\eqref{e:periodicity}
and~\eqref{e:fredholm}, 
the latter become
\bse
\begin{align}
\label{e:meq33}
&\partialderiv{G_1}{t} + 3\partialderiv{G_2}{x}
 { + \lambda \partialderiv{ }y \big( q~G_1+p\big)} = 0\,,
\\
\label{e:meq44}
&\partialderiv{G_2}{t} + \partialderiv{ }x\big( 4 G_3 - 3 k^2 G_4 \big)
  {  + \lambda\bigg( 2 G_2 \DDy q + 2 q \partialderiv{G_2}{y} - q^2 \partialderiv{G_2}{x}+2 G_1 \DDy p \bigg)} = 0\,,
\\
\label{e:mpconstraint1}
& { \partialderiv{ }x \big( q~G_1+p \big) - \partialderiv{G_1}{y} = 0\,,}
\end{align}
\ese
where
\begin{align}
G_1 = \int_{0}^{1} u^{(0)} ~d\theta\,, \quad 
G_2 = \int_{0}^{1} (u^{(0)})^2 ~d\theta\,,\quad 
G_3 = \int_{0}^{1} (u^{(0)})^3 ~d\theta\,, \quad 
G_4 = \int_{0}^{1} (u_\theta^{(0)})^2 ~d\theta\,
\end{align}
and where $Df/Dy$ was defined in~\eqref{e:DDy}.
Using~\eqref{e:ellipticsoln} and the properties of elliptic functions (see Byrd and Friedman~\cite{elliptichandbook}, formulas 312 and special values 122), we find
\bse
\label{e:integrals}
\begin{align}
G_1 &= \frac{V}{6} + \frac{\beta J}{3}\,,
\qquad
G_2 = \frac{V^2}{36} + \frac{\beta VJ}{9} + \frac{\beta^2}{9}\Delta\,, \\
G_3 & = \frac{V^3}{216} + \frac{\beta V^2J}{36} + \frac{\beta^2 V}{18}\Delta
  + \frac{\beta^3}{135} \bigg(\frac{27\E}{\K}\Delta + 5m^3 - 21m^2+33m -22\bigg)\,,
\\
G_4 &= \frac{16\beta^2 \K^2}{15} \bigg( \frac{2\E}{\K}\Delta  - m^2 + 3m -2 \bigg)\,,
\end{align}
where for brevity we introduced the shorthand notations
\begin{align}
\beta = \frac bm\,,
\qquad
\Delta = m^2 - m + 1\,,
\qquad
J = \frac{3 \E}{\K} + m -2\,.
\label{e:Jdef} 
\end{align}
\ese
Using~\eqref{e:qdef},~\eqref{e:omegadef} and~\eqref{e:abxidef}, the five modulation equations then become
\bse
\label{e:meqns}
\begin{align}
\label{e:228a}
&\partialderiv{ }t \bigg( \frac{1}{\K} \beta^{1/2} \bigg) + \partialderiv{ }x \bigg( \frac{V+\lambda q^2}{\K} \beta^{1/2} \bigg) = 0\,, 
\\
\label{e:228b}
&\partialderiv{ }t \bigg( \frac{q}{\K} \beta^{1/2} \bigg) + \partialderiv{ }y \bigg( \frac{V+\lambda q^2}{\K} \beta^{1/2} \bigg) = 0\,, 
\\
&\partialderiv{ }t \big( V + 2J\beta \big)
  + \partialderiv{ }x \bigg( \frac{V^2}{2} + 2VJ\beta + 2\Delta\beta^2\bigg) 
  { + \lambda  \partialderiv{ }y \big( q (V + 2J\beta) + 6 p\big)} = 0\,,
\\
&\partialderiv{ }t \big( V^2 + 4VJ\beta + 4\Delta\beta^2 \big) 
  + \partialderiv{ }x \bigg( \frac{2V^3}{3} + 4V^2J\beta + 8 V\Delta\beta^2 + \frac{8}{3}(m + 1)(m - 2)(2 m - 1)\beta^3\bigg) 
\nonumber\\
&\kern2em{ } + \lambda \bigg\{ \Big[2(1+q) \DDy q + q^2 \partialderiv{ }x\Big]
  \big[V^2 + 4VJ\beta + 4\Delta\beta^2\big] { + 12 \DDy p\, \big( V + 2J\beta\big)} \bigg\}  = 0\,, \\
&{ \partialderiv{ }x \big( q (V + 2J\beta) + 6 p\big) - \partialderiv{ }y \big( V + 2J\beta \big)  = 0\,.}
\end{align}
\ese

The system~\eqref{e:meqns} comprises {five} (2+1)-dimensional quasi-linear PDEs for the five dependent variables 
$V$, $\beta = b/m$, $m$, $q$ {and $p$,} 
which describe the slow modulations of the parameters of the cnoidal wave solution of the KP equation.
{These are the modulation equations in physical variables.}

\subsection{Transformation to Riemann-type variables}
\label{s:riemann}

Here we introduce convenient Riemann-type variables to reduce the system of PDEs~\eqref{e:meqns} into a simple form, 
following the procedure used by Whitham for the KdV equation~\cite{Whitham}. 
For the KdV equation the Whitham system of equations can actually be diagonalized exactly. 
Conversely, for the KP equation the system~\eqref{e:meqns} 
cannot be transformed into diagonal form using a similar change of dependent variables.
Nonetheless, the form of the system can be simplified considerably.

Importantly, if one sets {$q(x,y,0)=p(x,y,0) = 0$} and removes the $y$-dependence from the initial conditions for the remaining variables, the system~\eqref{e:meqns} reduces to three (1+1)-dimensional quasi-linear PDEs which are exactly the modulation equations for the KdV equation (cf.\ Appendix~\ref{s:riemannkdv}).
That is, the Whitham equations for the  KP equation derived from~\eqref{e:evol} contain those for the KdV equation as a special case
(see next section for further details). 
For this reason, we will introduce the same Riemann-type variables $r_1$, $r_2$ and $r_3$ by letting
\begin{align}
\label{e:riemannvariables}
V = 2(r_1 + r_2 + r_3)\,, \quad \frac{b}{m} = 2(r_3 - r_1)\,, \quad m = \frac{r_2 - r_1}{r_3 - r_1}\,.
\end{align}
The quantities $r_1,\,r_2,\,r_3$, which are easily obtained from $V$ and $b$ and $m$ by inverting~\eqref{e:riemannvariables},
are the so-called Riemann invariant variables for the KdV equation (see Appendix~\ref{s:riemannkdv})..
If the Riemann variables $r_1,\,r_2,\,r_3$ and $q$ are known, one can easily recover the solution of the KP equation. 
Indeed, using~\eqref{e:riemannvariables}, the cnoidal wave solution~\eqref{e:ellipticsoln} becomes
\begin{align}
\label{e:ellipticsoln2}
u^{(0)}(r_1,r_2,r_3,q) = r_1 - r_2 + r_3 + 2(r_2 - r_1)\cn^2\bigg(2\K(\theta - \theta_0), \frac{r_2-r_1}{r_3-r_1}\bigg)\,.
\end{align}
The rapidly varying phase $\theta$ can also be recovered (up to an integration constant) by integrating~\eqref{e:thetaderiv}.
Finally, the value of $p$ determines uniquely the auxiliary field $v$ via~\eqref{e:v0def}.
Therefore, up to a translation constant in the fast variable $\theta$,
there is a direct and one-to-one correspondence between the dependent variables $r_1,r_2,r_3,q,p$ 
in the Whitham modulation system and the leading-solution of KP equation.

Substituting~\eqref{e:riemannvariables} into the system of equations~\eqref{e:meqns} we obtain, in vector form
\begin{align}
\label{e:prematrix}
R \frac{\partial \textbf{r}}{\partial t} + S \frac{\partial \textbf{r}}{\partial x} +T \frac{\partial \textbf{r}}{\partial y} = 0\,,
\end{align}
where $\textbf{r} = (r_1,~r_2,~r_3,~q,~{p})^T$ and $R$, $S$ and $T$ are suitable real {$5\times5$} matrices.
In particular, $R$ has the block-diagonal structure
$R = \mathop{\rm diag}(R_4,0)$,
where $R_4$ is a $4\times4$ matrix.
Even though $R$ is not invertible, we can multiply~\eqref{e:prematrix} by the "pseudo-inverse"
$\~R^{-1} = \mathop{\rm diag}(R_4^{-1},0)$,
obtaining
\begin{align}
\label{e:matrixeqn}
{I}\, \frac{\partial \textbf{r}}{\partial t} + A \frac{\partial \textbf{r}}{\partial x} +B \frac{\partial \textbf{r}}{\partial y} = 0\,,
\end{align}
where {$I = \mathrm{diag}(1,1,1,1,0)$\,, and with} $A = R^{-1}S$ and $B = R^{-1}T$.
The entries of the matrices $A$ and $B$ (which were calculated using \textit{Mathematica}) are given 
explicitly in Appendix~\ref{s:matrix}.

As mentioned before, unlike the case of the Whitham equations for the KdV equation, the matrices $A$ and $B$ are not diagonal. 
Moreover, in order for the above system to be diagonalizable, the matrices $A$ and $B$ would need to be simultaneously diagonalizable,
which is possible only if they commute.
It is easy to check, however, that $AB \neq BA$.  
Therefore, one cannot write the Whitham system for the KP equation in diagonal form 
using a change of dependent variables.

\subsection{Singularities of the original modulation system and their removal}
\label{s:simplify}

The Whitham system~\eqref{e:matrixeqn} becomes singular in certain limits.
Here we characterize this singular behavior and show how one can use the {third compatibility condition~\eqref{e:constraint} and the constraint~\eqref{e:mpconstraint1}} to eliminate the singularities, resulting in a modified Whitham system that is singularity-free.

\paragraph{Singularities of the original Whitham system.}
%
Let us study the limiting behavior of the modulation equations as the elliptic parameter $m$ tends to 0 or 1. 
Recall that $\cn(x,m)\to\sech(x)$ as $m\to1$, and the cnoidal wave solution~\eqref{e:ellipticsoln2} becomes the line soliton solution of the KP equation in this limit:
$u^{(0)}(x,y,t) = r_1 + 2(r_2 - r_1)\sech^2[\sqrt{r_2-r_1}(x+q y-{(V+\lambda q^2)} t)]$
{[note $r_3=r_2$ when $m=1$, cf.~\eqref{e:riemannvariables}]}.
Conversely, when $m\ll1$, the cnoidal wave solution
$u^{(0)}(x,y,t)$ reduces to a sinusoidal function which has a vanishingly small amplitude in the limit $m\to0$.

The Whitham system~\eqref{e:matrixeqn}
{becomes singular in both limits.
That is, some of the entries of both matrices $A$ and $B$}
have an infinite limit as $m\to 0$ and $m\to 1$, 
even though the determinants and eigenvalues of $A$ and $B$ remain finite. 
The same problem arises for the matrices $R$, $S$ and $T$ in~\eqref{e:prematrix} as well.
Moreover, the singularity is also present in the original system~\eqref{e:meqns}.  
That is, writing~\eqref{e:meqns} as a system of PDEs for the dependent vector variable $\textbf{w} = (V,~b/m,~m,~q,~p)^T$, 
all of the resulting coefficient matrices have infinite limits as {$m\to 0$ and $m\to 1$}.

This singular behavior does not occur in the (1+1)-dimensional case. 
For the KdV equation, even though the corresponding $3\times3$ matrices 
$R$ and $S$ have infinite limits as $m\to 1$, once the system is converted into a diagonal form, 
the limits of the velocities $V_1$, $V_2$ and $V_3$ (which are the entries of the resulting diagonal matrix) are finite. 
In other words, the diagonalization of the system eliminates the singular limit since the eigenvalues of all the matrices have finite limit. 
In fact, for the Riemann problem for the KdV equation, the limits of the velocity $V_2$ as $m\to 0$ and $m\to 1$ yield the velocities of the leading and trailing edges of the DSW, respectively~\cite{GP1974}.
Similarly, for the cylindrical KdV equation, the Whitham system is inhomogeneous~\cite{ckdv},
but the velocities $V_1$, $V_2$ and $V_3$ have the same form as those for the KdV equation, and the inhomogeneous terms
also have finite limits when $m \to 0$ and $m \to 1$.

In both of the above (1+1)-dimensional cases, all relevant $3 \times 3$ matrices have finite non-zero determinants. 
For the KP equation, however, since the first and second modulation equations [namely, \eqref{e:eq1} and \eqref{e:eq2}] do not contain derivatives with respect to $x$ and $y$, respectively, the second row of the matrix $S$ and the first row of the matrix $T$ in~\eqref{e:prematrix} are identically zero, which make their determinants zero. 
Since $A$ and $B$ cannot be diagonalized simultaneously, one needs to deal with this singular limit another way.

\paragraph{Removal of the singularities and final KP-Whitham system.}
To simplify the Whitham system, we make use of the compatibility {condition~\eqref{e:constraint} and the constraint~\eqref{e:mpconstraint1}}.
Using~\eqref{e:abxidef} and~\eqref{e:riemannvariables}, the slowly varying variable $k$ can be written in terms of the Riemann-type variables as
\begin{align}
k = \frac{1}{2\sqrt{2}\K} \sqrt{\frac{b}{m}} = \frac{1}{2 \K} \sqrt{r_3 - r_1}\,.
\label{e:kdef}
\end{align}
Correspondingly, recalling that $q = l/k$, \eqref{e:constraint} becomes
\begin{align}
\partialderiv{ }y \bigg(\frac{1}{2 \K} \sqrt{r_3 - r_1}\bigg) - \partialderiv{ }x \bigg(\frac{q}{2 \K} \sqrt{r_3 - r_1}\bigg) =0\,,
\label{e:constraint0}
\end{align}
or equivalently,
\begin{align}
\label{e:constr}
- q \bigg(b_1 \partialderiv{r_1}{x} + b_2 \partialderiv{r_2}{x} +b_3 \partialderiv{r_3}{x}\bigg)
 -\frac{2(r_3 - r_1)}{\K}\partialderiv{q}{x}
 + \bigg(b_1 \partialderiv{r_1}{y} + b_2 \partialderiv{r_2}{y} + b_3 \partialderiv{r_3}{y}\bigg) =0\,,
\end{align}
where 
\begin{align*}
b_1 = \frac{\E-\K}{m\K^2}\,, \quad
b_2 = -\frac{\E-(1-m)\K}{m (1-m)\K^2}\,, \quad
b_3 = \frac{\E}{(1-m)\K^2}\,.
\end{align*}
Note that {\eqref{e:constr} above} is identically satisfied when $q$ is identically zero and $r_1,r_2,r_3$ are independent of $y$.

Although $b_2$ and $b_3$ in~\eqref{e:constr} also have infinite limits when $m\to 1$, 
we can use~\eqref{e:constr} to simplify the Whitham system~\eqref{e:matrixeqn}.
Indeed, by subtracting a suitable multiple of the {equation}~\eqref{e:constr} from each equations {and a suitable multiple of the constraint~\eqref{e:mpconstraint1} from the first three equations of~\eqref{e:matrixeqn}}
(see Appendix~\ref{s:matrix} for details),
the modulation equations take on the particularly simple form~\eqref{e:kpwhitham},
where  
\bse
\label{e:kpwhithamcoeffs}
\begin{align}
V_1 = V - 2b \frac{\K}{\K - \E}\,,\qquad
V_2 = V - 2 b \frac{(1-m) \K}{\E - (1-m) \K}\,,\qquad 
V_3 = V  + 2b \frac{(1-m) \K}{m\E}\,,
\label{e:Vdef}
\end{align}
with $V = 2(r_1 + r_2 + r_3)$ and $b = 2(r_2-r_1)$ as for the KdV equation, and
\begin{gather}
\label{e:nudef}
\nu_1 = \frac{V}{6} + \frac{b}{3 m} \frac{(1+m)\E-\K}{\K-\E}\,, \qquad
\nu_2 = \frac{V}{6} + \frac{b}{3 m} \frac{(1-m)^2\K - (1-2 m)\E}{\E - (1-m)\K}\,, \\
\nu_3 = \frac{V}{6} + \frac{b}{3 m} \frac{(2-m)\E -(1-m)\K}{\E}\,, \qquad
\nu_4 = \frac{2m\E}{\E - (1-m) \K}\,, \\
\nu_{4.1} = 4-\nu_4\,, \qquad \nu_{4.3} = 2+\nu_4\,, \qquad 
 \nu_5 = r_1 - r_2 + r_3\,, \qquad
{\alpha = \frac{\E}{\K}\,.}
\end{gather}
\ese
The fact that one of the equations in~\eqref{e:kpwhitham} is not in evolution form should not be surprising in light of the 
non-local nature of the KP equation itself.
Importantly, the system~\eqref{e:kpwhitham} 
is completely free of singularities.  
That is, all the coefficients have finite limits as $m \to 0$ and $m \to 1$. 
The speeds $V_1,\dots,V_3$ are exactly the characteristic speeds of the KdV-Whitham system (cf.\ section~\ref{s:properties} and Appendix~\ref{s:riemannkdv}).
Also, $\nu_1,\dots,\nu_3$ are exactly the same as the coefficients appearing in the inhomogeneous terms for the cKdV-Whitham system (cf.\ section~\ref{s:properties} and \cite{ckdv}).
Finally, note that, even though the dependent variable $p(x,y,t)$ does not directly affect the leading-order cnoidal wave solution~\eqref{e:ellipticsoln},
including its dynamics modulations in the KP-Whitham system~\eqref{e:kpwhitham}
ensures that the system preserves all of the symmetries of the KP equation (cf.\ section~\ref{s:properties})
and that the the stability properties of the solutions are consistent with those of the KP equations (cf.\ section~\ref{s:stability}).

The relatively simple form of the KP-Whitham system~\eqref{e:kpwhitham}
and the fact that all coefficients remain finite for all values of $m$
make it possible to find several exact reductions (cf.\ section~\ref{s:properties})
and to use it to study the behavior of solutions of the KPI and KPII equations (cf.\ section~\ref{s:stability}). 
to study the behavior of DSW for the KP equation.

\paragraph{Remarks.}
%
Krichever \cite{krichever}
used the Lax pair of the KP equation and the finite-genus machinery to formulate a general methodology
to derive the genus-$N$ modulation equations for arbitrary $N$.
While the theory is elegant, the modulation equations are only given in implicit form.
{While our derivation is limited to the genus-1 case, it does not require or use integrability, 
and hence it can be used to study certain non-integrable problems.  Also,}
the dependent variables {in our work} have a clear physical interpretation, and the properties of the equations 
{as well as the connection to (2+1)-dimensional DSWs}
are discussed in detail.

Bogaevskii~\cite{Bogaevskii} used the method of averaging (as opposed to direct perturbation theory), and obtained 
six modulation equations. 
One of them [the last equation of~(4.1a)] is the constraint $k_x=l_y$ for the phase.
The other five include four evolution PDEs and one additional constraint [the last equation of (4.1b)].
The key differences between the system in~\cite{Bogaevskii} and~\eqref{e:kpwhitham} are 
on one hand that the system in \cite{Bogaevskii} is not written in terms of the Riemann-like variables,
and on the other hand that the role of the auxiliary variables $\alpha$ and $\beta$ in \cite{Bogaevskii} is not explained.
For example, even the reduction to the KdV equation is not entirely trivial.

Infeld and Rowlands \cite{infeld} used the Lagrangian approach to Whitham theory and derive five PDEs, 
of which three are evolution form while the remaining two are constraints.
Notably, however, the leading-order solution of the KP equation is not written explicitly as a cnoidal wave.
Correspondingly, 
some of the dependent variables arise as integration constants [e.g., in~(8.3.12)], 
whose physical meaning is not immediately clear.
Moreover, the modulation equations are given in implicit form, because they involve partial derivatives of the quantity $W(A,B,U)$ 
[defined in~(8.3.18)] which is not explicitly computed.
 
In conclusion, 
while in theory it is possible that the system of modulation equations derived here
is equivalent to some or even all of the the systems in the above works,
in practice showing the equivalence of any two of the above systems is a nontrivial task, 
which is outside the scope of this work.
Moreover, here we have transformed the system of modulation equations into the singularity-free system in Riemann-type variables~\eqref{e:kpwhitham} 
as given in~\eqref{e:kpwhitham}, 
where connections to important reductions, (2+1)-dimensional DSWs and stability can be more easily carried out. 
This was not done in any of the above references.

In the remaining part of this work: 
(i) we discuss in detail various properties of the KP-Whitham system~\eqref{e:kpwhitham}, 
including symmetries and several exact reductions, 
(ii) we discuss the (2+1)-dimensional generalization of the Riemann problem for the KdV equation, and 
(iii) we show how the KP-Whitham system~\eqref{e:kpwhitham} can be used to obtain concrete answers about the stability of the solutions of the KP equation,
all of which are novel to the best of our knowledge.

\section{Properties of the KP-Whitham system}
\label{s:properties}

\subsection{Symmetries of the KP-Whitham system}
\label{s:symmetries}

Here we discuss how the invariances of the KP equation are reflected in corresponding invariances for the KP-Whitham system~\eqref{e:kpwhitham}. 
It is well known that the KP equation admits the follwing symmetries:
\begin{align*}
& u(x,y,t) \mapsto u(x-x_0,y-y_0,t-t_0)\,,  & \textrm{(space/time translations)} \\
& u(x,y,t) \mapsto a + u(x-6 a t,y,t)\,,  & \textrm{(Galilean)} \\
& u(x,y,t) \mapsto a^2 u(a x, a^2 y, a^3 t)\,, &  \textrm{(scaling)} \\
& u(x,y,t) \mapsto u(x+a y-\lambda a^2 t, y- 2 \lambda a t, t)\,,  & \textrm{(pseudo-rotations)}
\end{align*}
{with $a$ an arbitrary real constant.}	
Namely, if $u(x,y,t)$ is any solution of the KP equation, the transformed field is a solution as well.
Each of these symmetries
generates a corresponding symmetry for the KP-Whitham system~\eqref{e:kpwhitham}. 
The invariance under space/time translations is trivial. 
For the other invariances, the corresponding transformations for the Riemann variables can be derived as follows:

\noindent
Galilean transformations:
\begin{align*}
&r_j(x,y,t) \mapsto a + r_j(x - 6 a t, y, t), \quad j=1,2,3, 
\\
&q(x,y,t) \mapsto q(x - 6 a t, y, t)\,,  
\\
&{p(x,y,t) \mapsto p(x - 6 a t, y, t) - a\,q(x - 6 a t, y, t)\,,}  
\\
\noalign{\noindent scaling transformations:}
&r_j(x,y,t) \mapsto a^2 r_j(a x, a^2 y, a^3 t), \quad j=1,2,3, 
\\
&q(x,y,t) \mapsto a q(a x, a^2 y, a^3 t)\,, 
\\
&{p(x,y,t) \mapsto a^3 p(a x, a^2 y, a^3 t)\,,}
\\
\noalign{\noindent pseudo-rotations:}
&r_j(x,y,t) \mapsto r_j(x+a y-\lambda a^2 t, y- 2 \lambda a t, t), \quad j=1,2,3,
\\
&q(x,y,t) \mapsto a +  q(x+a y-\lambda a^2 t, y- 2 \lambda a t, t)\,,
\\
&{p(x,y,t) \mapsto p(x+a y-\lambda a^2 t, y- 2 \lambda a t, t)\,.}
\end{align*}
It is straightforward to verify that 
{all these transformations} leave the KP-Whitham system~\eqref{e:kpwhitham} invariant.  
For brevity we omit the details.

{Finally, recall the KP equation is invariant under the transformation $v(x,y,t)\mapsto a + v(x,y,t)$.
This symmetry is reflected in the corresponding symmetry of the KP-Whitham system under the transformation
$p(x,y,t) \mapsto a + p(x,y,t)$.
In other words, adding an arbitrary constant offset to $p(x,y,t)$ leaves~\eqref{e:kpwhitham} invariant.}


\paragraph{Time invariance of the constraint.}
\label{s:t-invariance}
Next we show that for the KP-Whitham system~\eqref{e:kpwhitham}, the constraint~\eqref{e:constraint} 
(namely, $k_y = l_x$)
is invariant with respect to time.
The definitions~\eqref{e:thetaderiv} imply that~\eqref{e:constraint} is automatically satisfied if the solution of the KP-Whitham system~\eqref{e:kpwhitham} is obtained from a modulated cnoidal wave of the KP equation and $\theta(x,y,t)$ is smooth.
Here, however, we show that the constraint is time-invariant
\textit{independently of whether the dependent variables for the system~\eqref{e:kpwhitham} originate from a solution of the KP equation}.

In other words, consider arbitrarily chosen ICs for the dependent variables $r_1,r_2,r_3$, $q$ {and $p$},
and recall that $k$ is determined from $r_1,r_2,r_3$ via~\eqref{e:riemannvariables} and \eqref{e:kdef}.
Also recall that $q = l/k$, and let $f(x,y,t) = k_y - (kq)_x$.
The constraint $k_y = l_x$ is equivalent to the condition $f(x,y,t) = 0$ $\forall t\ge0$.
But we next show that, if the ICs for~\eqref{e:kpwhitham} are such that $f(x,y,0) = 0$, 
then $f(x,y,t) = 0$ $\forall t>0$.

To prove this, we first note that the original system of modulation equations~\eqref{e:meqns} immediately implies $\partial f/\partial t = 0$,
since the first and second equations in~\eqref{e:meqns} are simply~\eqref{e:eq1} and~\eqref{e:eq2}
(i.e., $k_t + \omega_x =0$ and $l_t + \omega_y = 0$), respectively.
The same result holds true for the un-regularized Whitham system~\eqref{e:matrixeqn} for the variables $r_1,r_2,r_3,p$ and $q$,
since~\eqref{e:meqns} and~\eqref{e:matrixeqn} are equivalent.
Also, the constraint~\eqref{e:constraint} (i.e., $f=0$) becomes~\eqref{e:constr} when written in terms of these variables.

The situation is  different for the regularized KP-Whitham system~\eqref{e:kpwhitham}, however, 
because~\eqref{e:kpwhitham} is obtained from~\eqref{e:matrixeqn} precisely by subtracting a suitable multiple of the {third compatibility condition}~\eqref{e:constr}.
Nonetheless, tedious but straightforward algebra shows that \eqref{e:kpwhitham} yields a linear homogeneous first-order ordinary differential equation for $f$ (i.e., $\partial f/\partial t = \mu\,f$, with $\mu$ a scalar function).
Therefore, if $f$ vanishes at $t=0$, it will remain zero at all times.

Importantly, this result can be used to determine ICs for the variable $q(x,y,0)$ once one has determined the ICs for $r_1,r_2,r_3$.
(See section~\ref{s:riemannproblem} for further details.)

\subsection{Exact reductions of the KP-Whitham system}
\label{s:reductions}


\paragraph{KdV reduction.}
Every solution of the KdV equation~\eqref{e:kdv} is obviously also a $y$-independent solution of the KP equation. 
One of the advantages of using the form~\eqref{e:evol} of the KP equation as opposed to the standard form~\eqref{e:kp}
is that, if one takes $\lambda=0$, it immediately reduces exactly to the KdV equation 
(as opposed to the $x$ derivative of it, as it happens for the KP equation in standard form). 
Correspondingly, letting $\lambda=0$, the KP-Whitham system~\eqref{e:kpwhitham} reduces to the Whitham modulation equations for the KdV equation in diagonal form
\vspace*{-0.8ex}
\begin{align}
\partialderiv{r_i}{t} + V_i \partialderiv{r_i}{x} = 0\,, \qquad  i = 1\,,2\,,3\,,
\label{e:whithamkdv}
\end{align}
where $V_1,V_2$ and $V_3$ are given by~\eqref{e:Vdef}
together with a PDE for the fourth variable $q(x,y,t)$
\begin{align}
\partialderiv{q}{t} + V_2 \,\partialderiv{q}{x}
  + \nu_{4.1} \DDy{r_1} + \nu_{4.3} \DDy{r_3} = 0\,, 
\label{e:qeqn}
\end{align}
and the constraint~\eqref{e:pcontr}.
(Note however that the system~\eqref{e:whithamkdv} is independent of $q$, whose value is only needed if one
wants to recover the solution of the KP equation from that of the KP-Whitham system.)
If we now choose the initial conditions $r_j(x,y,0)$ $(j=1,2,3)$ to be independent of $y$ and $q(x,y,0) = 0$ {with $p(x,y,0)$ a constant, then~\eqref{e:pcontr} is automatically satisfied and}
these conditions remain true for all time, that is, $r_j$ $(j=1,2,3)$ are also independent of $y$ for all $t>0$ and $q=0$.

Note also that it is not necessary to take $\lambda=0$ to obtain the KdV reduction.
Indeed, it is straightforward to see that, if one takes $y$-independent ICs for $r_1,r_2,r_3$ and $q(x,y,0) = 0$ {with $p(x,y,0)$ a constant},
one has $q(x,y,t)=0$ for all $t$, 
and the solution of the KP-Whitham system~\eqref{e:kpwhitham} coincides with that of the corresponding system for the KdV equation.

\paragraph{``Slanted'' KdV reduction.}
The KP-Whitham system~\eqref{e:kpwhitham} also admits a ``slanted'' KdV reduction.
Suppose that {$q$ and $p$ are constants} and the three Riemann variables depend on $x$ and $y$ only through the similarity variable $\xi = x+q y$, that is, $r_j = r_j (\xi,t) \quad j = 1,2,3\,.$
Then we have 
\begin{align*}
\DDy{r_j} = \partialderiv{r_j}{y} - q \partialderiv{r_j}{x} 
= q \frac{\partial r_j}{\partial \xi} - q \frac{\partial r_j}{\partial \xi}= 0\,.
\end{align*}
Correspondingly, the KP-Whitham system~\eqref{e:kpwhitham} reduces to a diagonal system
\begin{align}
\partialderiv{r_j}{t} + V_j \frac{\partial r_j}{\partial \xi} = 0\,, \quad j = 1,2,3\,.
\label{e:slant}
\end{align}
We can also prove a stronger result. 
Namely, if $q(x,y,0)$ {and $p(x,y,0)$ are} constants and $r_j(x,y,0)$ depend on $x$ and $y$ only through the similarity variable $\xi = x + qy$
[i.e., $r_j(x,y,0) = r_j(\xi,0)$], then the time evolution of those Riemann variables will be determined by the reduced system~\eqref{e:slant} and as a result the conclusion will remain true for any time $t$. 
That is, $r_j(x,y,t) = r_j(\xi,t)$ and $q(x,y,t) = q(x,y,0)$, which is a constant.

\paragraph{Cylindrical KdV reduction.}
The KP-Whitham system~\eqref{e:kpwhitham} can also be reduced to the modulation equations for the cKdV equation~\cite{ckdv}. 
We next discus this reduction and recover the previous two reductions as special cases.

Let $r_j(x,y,t)$ $(j=1,2,3)$ depend on $x$ and $y$ only through the similarity variable $\eta = x+P(y,t)$, that is, $r_j = r_j (\eta,t)$ for $j=1,2,3$ and 
\begin{equation}
\label{e:qPcKdVdef}
q(x,y,t) = P_y(y,t)\,, 
\end{equation}
{with $p(x,y,t)=$const}. 
Then we have
\begin{align*}
\partialderiv{r_j}{t}  = \partialderiv{r_j}{t} + P_t \frac{\partial r_j}{\partial \eta}\,, \quad
\partialderiv{r_j}{x}  = \frac{\partial r_j}{\partial \eta}\,, \quad
\partialderiv{r_j}{y}  = P_y \frac{\partial r_j}{\partial \eta}\,,
\end{align*}
implying $D r_j/D y = 0$.
Since $q(x,y,t)$ is independent of~$x$, the first three equations in the KP-Whitham system~\eqref{e:kpwhitham} simplify to
\begin{align}
\label{e:first3eqn}
\partialderiv{r_j}{t} + P_t \frac{\partial r_j}{\partial \eta}+ (V_j + \lambda P_y^2)\,\frac{\partial r_j}{\partial \eta}
+ \lambda \nu_j P_{yy} = 0\,, \qquad j = 1,2,3\,,
\end{align}
while the fourth equation becomes
\bse
\label{e:ckdvhydrodynamicsystem}
\begin{align}
\label{e:qHopf}
\partialderiv{q}{t} + 2 \lambda q \partialderiv{q}{y} = 0\,.
\end{align}
{(Note that in this case the constraint~\eqref{e:pcontr} is again automatically satisfied.)}
Using \eqref{e:qPcKdVdef}, \eqref{e:qHopf} becomes $P_{ty} + 2 \lambda P_y P_{yy} = 0$, which after integration yields
$P_t + \lambda P_y^2 = 0$.
(Taking into account integration constants would add an arbitrary function of time in the right-hand side (RHS) of the above relation.
The presence of such a function would in turn result in an additional term 
to the definition of $\eta$, but would not change the structure of the equations.
For simplicity, we set this integration constant to zero in the discussion that follows.)
Moreover, using the above relation, the system of equations~\eqref{e:first3eqn} becomes
\begin{align}
\label{e:ckdvreduction}
\partialderiv{r_j}{t} + V_j\,\frac{\partial r_j}{\partial \eta} + \nu_j \partialderiv qy = 0\,, \qquad j = 1,2,3\,.
\end{align}
\ese
In order for this setting to be self-consistent, however, the last term in the LHS of~\eqref{e:ckdvreduction} must be independent of $y$.
Therefore, only three possibilities arise:
(i) $P_y = 0$, in which case one simply has $q(x,y,t)=0$ (implying that the resulting behavior is one-dimensional)
and $P(y,t) = 0$ as well as $\eta = x$, and the system~\eqref{e:ckdvreduction} reduces to the Whitham system for the KdV equation.
(ii) $P_y = a$ is a constant, then one has $P(y,t) = ay$
implying $\eta = x + ay$, in which case the system~\eqref{e:ckdvreduction} reduces to the Whitham system for the ``slanted'' KdV reduction.
(iii) $P_{yy} = f(t)$ is a function of $t$,  in which case $q = P_y = f(t) y$ (again neglecting trivial integration constants).
Note also that \eqref{e:qHopf} is the Hopf equation.
{Thus, if $q(y,0)=cy$, with $c={}$const, \eqref{e:qHopf} can be integrated by characteristics to yield}
\begin{align} 
q(y,t) = \frac{cy}{1+2c\lambda t}\,,
\end{align} 
implying $f(t) = c/(1+2c\lambda t)$ and 
$P(y,t) = {cy^2}/[{2(1+2c\lambda t)}]$,
which reduces~\eqref{e:ckdvreduction} to the Whitham system for the cKdV equation \cite{ckdv}.
%

Of course, similarly to the KdV and ``slanted" KdV cases, one could also prove a stronger result. 
Namely, if the initial conditions $r_j(x,y,0)$ $(j=1,2,3)$ depend on $x$ and $y$ only through the similarity variable $\eta$, that is
$r_j(x,y,0) = r_j(x + P(y,0))$ for $j=1,2,3$
and $q(x,y,0) = P(y,0)$, this dependence will be preserved for all time. More precisely, we will have
$r_j(x,y,t) = r_j(x + P(y,t))$ for $j=1,2,3$ and 
$q(x,y,t) = P(y,t)$.

\paragraph{Reduction $p={}$const.}
In all three reductions considered above, 
the requirement that $p(x,y,t)$ be constant was one of the assumptions. 
Next we next discuss the reduction of the KP-Whitham system when $p(x,y,t)={}$const is the only condition being imposed.
In this case, the first four of~\eqref{e:kpwhitham} yield the following $4\times4$ hydrodynamic system in two spatial dimensions:
\bse
\label{e:preduction}
\begin{align}
&\partialderiv{r_j}{t} + (V_j + \lambda q^2)\,\partialderiv{r_j}{x} + 2 \lambda q  \DDy{r_j} + \lambda \nu_j \DDy{q} = 0\,, \qquad j=1,2,3,\\
&\partialderiv{q}{t} + (V_2 + \lambda q^2)\,\partialderiv{q}{x} + 2 \lambda q \DDy{q} 
+ \nu_{4.1} \DDy{r_1} + \nu_{4.3} \DDy{r_3} = 0\,,
\end{align}
\ese
for the four dependent variables $r_1,r_2,r_3$ and $q$.
Note however that the last of~\eqref{e:kpwhitham} yields the additional equation
\begin{equation}
\nu_5 \partialderiv{q}{x} = (1-\alpha) \DDy{r_1} + \alpha \DDy{r_3}\,,
\label{e:pconstraint}
\end{equation}
which imposes a constraint on the values of $r_1,r_3$ and $q$. 
The above reduction [and the system~\eqref{e:preduction}] are therefore only consistent if the constraint~\eqref{e:pconstraint} 
is satisfied for all $t\ge0$.
This is indeed the case for the KdV, slanted KdV and cKdV reductions.
However, it is unclear at present whether other reductions of the KP-Whitham system to a self-consistent $4\times4$ system exist.

\paragraph{Genus-zero reductions.}
The system~\eqref{e:kpwhitham} admits two further exact reductions, which are obtained respectively when $r_1 = r_2$ and $r_2=r_3$.
The first one corresponds to the case in which the leading-order cnoidal wave solution degenerates to a constant with respect to the fast variable,
and the second one to the solitonic limit.
We next discuss these two reductions separately.

When $r_1=r_2$, one has $m=0$. Then $\E = \K = \pi/2$, and all the coefficients of the KP-Whitham system simplify considerably.
Moreover, the PDEs for $r_1$ and $r_2$ coincide in this case.
As a result, \eqref{e:kpwhitham} reduces to the following $4\times4$ system:
\bse
\label{e:kpwhitham1reduction}
\begin{align}
&\partialderiv{r_1}{t} + (12 r_1 - 6 r_3 + \lambda q^2)\,\partialderiv{r_1}{x} + 2 \lambda q  \DDy{r_1} + \lambda r_3 \DDy{q} + \lambda \DDy{p} = 0\,,
\\
&\partialderiv{r_3}{t} + (6 r_3 + \lambda q^2)\,\partialderiv{r_3}{x} + 2 \lambda q  \DDy{r_3} + \lambda r_3 \DDy{q} + \lambda \DDy{p} = 0\,,
\\
&\partialderiv{q}{t} + (12 r_1 - 6 r_3 + \lambda q^2)\,\partialderiv{q}{x} + 2 \lambda q \DDy{q} 
 + 6 \DDy{r_3} = 0\,, 
\\
&\partialderiv {p}{x} - \DDy{r_3} + r_3 \partialderiv {q}x = 0\,.
\end{align}
\ese

Similarly, when $r_2=r_3$, one has $m=1$.  Then $\E = 1$ and $\K\to\infty$ in this limit.  
As a result, the PDEs for $r_2$ and $r_3$ coincide, and 
\eqref{e:kpwhitham} reduces to the system
\bse
\label{e:kpwhitham2reduction}
\begin{align}
&\partialderiv{r_1}{t} + (6 r_1 + \lambda q^2)\,\partialderiv{r_1}{x} + 2 \lambda q  \DDy{r_1} + \lambda r_1 \DDy{q} + \lambda \DDy{p} = 0\,,
\\
&\partialderiv{r_3}{t} + (2 r_1 + 4 r_3 + \lambda q^2)\,\partialderiv{r_3}{x} + 2 \lambda q  \DDy{r_3} + \lambda \frac{4 r_3 - r_1}{3} \DDy{q} + \lambda \DDy{p} = 0\,,
\\
&\partialderiv{q}{t} + (2 r_1 + 4 r_3 + \lambda q^2)\,\partialderiv{q}{x} + 2 \lambda q \DDy{q} 
 + 2 \DDy{r_1} + 4 \DDy{r_3} = 0\,, 
\\
&\partialderiv {p}{x} - \DDy{r_1} + r_1 \partialderiv {q}x = 0\,.
\end{align}
\ese
As we show in the following section, 
both of these two reduced systems are useful in formulating well-posed problems for the full KP-Whitham system~\eqref{e:kpwhitham}.

\subsection{{Initial-value problems} for the KP-Whitham system}
\label{s:riemannproblem}

Here we briefly discuss 
{the formulation of initial value problems for the KP-Whitham system~\eqref{e:kpwhitham}, 
including appropriate ICs and boundary conditions (BCs) and, as a special case,}
the (2+1)-dimensional generalization of the Riemann problem for the KdV equation.

\paragraph{ICs for the KP-Whitham system.}
The problem of determining ICs for the Riemann-type variables $r_1,r_2,r_3$ from an IC for $u$ is a non-trivial one in general, but 
is exactly the same as in the one-dimensional case.
If this step can be completed,
one can determine the IC for the {fourth} variable, namely $q(x,y,0)$, 
using the constraint~\eqref{e:constraint} at $t=0$, 
obtaining 
\begin{align}
k(x,y,0)_y = [k(x,y,0) q(x,y,0)]_x\,.
\label{e:constraintIC}
\end{align}
Integrating~\eqref{e:constraintIC} with respect to $x$ and dividing by $k$, 
we then obtain
\begin{align}
q(x,y,0) = \frac{1}{k(x,y,0)} \bigg( q(x_0,y,0) k(x_0,y,0) + \int_{x_0}^x k_y(\xi,y,0) d \xi \bigg)\,,
\label{e:q-ic}
\end{align}
{where $k(x,y,0)$ is assumed to be non-zero.}

To determine the ICs for the fifth dependent variable, 
note that integrating~\eqref{e:pcontr} determines $p(x,y,t)$ for all $t\ge0$ up to an arbitrary function of $y$ and $t$:
\begin{equation}
p(x,y,t) = p^-(y,t) + \partial_x^{-1} \bigg[ 
   (1-\alpha)\DDy{r_1} + \alpha\DDy{r_3} - \nu_5 \partialderiv{q}{x} 
  \bigg]\,, 
\label{e:pintegrated}
\end{equation}
where the operator $\partial_x^{-1}$ is defined as
\begin{equation}
\label{e:dxinverse}
\partial_x^{-1}[f] = \int_{-\infty}^x f(\xi,y,t)\,\d\xi\,. 
\end{equation}
Obviously one can also evaluate~\eqref{e:pintegrated} at $t=0$.
Thus, the problem is reduced to the choice of suitable BCs, to which we turn next.

\paragraph{BCs for the KP-Whitham system.}
To complete the formulation of a well-posed initial value problem for the KP-Whitham system (as would be necessary, for example,
in order to perform a numerical study of the problem), one also needs to determine appropriate 
BCs for the KP-Whitham system~\eqref{e:kpwhitham}.
For the Riemann problem for the KdV equations (namely, for the PDEs~\eqref{e:whithamkdv}), the asymptotic values of $r_1,r_2,r_3$ as $x\to\pm\infty$ are constant (i.e., independent of $t$).
Already in the Riemann problem for the cylindrical KdV equation (namely for the PDEs~\eqref{e:ckdvreduction}), however, this is not the case anymore (e.g., see \cite{ckdv}).
In that case, the boundary values for $r_j$ can be obtained from~\eqref{e:ckdvreduction}.
Namely, it is easy to see that, if $\partial r_j/\partial \eta\to0$ as $\eta\to\pm\infty$,
\eqref{e:ckdvreduction} reduces to three ODEs for the time evolution of 
the limiting values $r_{j,\pm}(t) = \lim_{\eta\to\pm\infty}r_j(\eta,t)$.
The difference between the Riemann problem for cKdV and that for the full KP-Whitham system is that, for the latter, 
the boundary values of the Riemann invariants may in general also depend on the independent variable $y$.
On the other hand, 
if $\partial r_j/\partial x\to0$ 
and $\partial q/\partial x\to0$ as $x\to\pm\infty$,
\eqref{e:kpwhitham} reduces to a system of four (1+1)-dimensional PDEs which can be solved (either analytically or numerically) 
to obtain the boundary values $r_{j,\pm}(y,t)$ and $q_\pm(y,t)$.

To make the above discussion more precise, we need to first go back to the KP equation.
Integrating~\eqref{e:kp2} yields
\begin{equation}
v(x,y,t) = v^-(y,t) + \partial_x^{-1} [ u_y ]\,,
\label{e:vintegrated}
\end{equation}
where throughout this section we will use the superscript ``$-$'' to indicate the limiting value of 
each quantity as $x\to-\infty$, 
and the operator $\partial_x^{-1}$ is defined by~\eqref{e:dxinverse} as before.
Substituting~\eqref{e:vintegrated} into~\eqref{e:kp1} yields 
\begin{equation}
u_t + 6uu_x + \epsilon^2 u_{xxx} + \lambda\,\partial_x^{-1}[u_{yy}] + \lambda\,\partial_y v^- = 0\,.
\label{e:kpevol}
\end{equation}
Taking the limit of~\eqref{e:kpevol} as $x\to-\infty$ we then see immediately that, if one is interested in solutions~$u$ 
which tend to constant values as $x\to-\infty$ (i.e., $u^-$ independent of $t$),
one needs $\partial_y v^-(y,t) = 0$.
{Ignoring an unnecessary function of time, we therefore take $v^-(y,t) = 0$.}

Similar arguments carry over to the KP-Whitham system~\eqref{e:kpwhitham}.
More precisely, 
recalling the cnoidal-wave representation~\eqref{e:ellipticsoln} of the leading-order solution $u^{(0)}$
of the KP equation 
as well as the representation~\eqref{e:riemannvariables} of the elliptic parameter $m$ in terms of the Riemann invariants,
we see immediately that, 
in order to ensure that $u$ tends to a constant as $x\to-\infty$,
one needs either $m^-=0$ or $m^-=1$,
i.e., either $r_1^- = r_2^-$ or 
$r_2^- = r_3^-$, respectively.
This is exactly the same as for the KdV equation.  
Also, recalling~\eqref{e:v0def}, and enforcing $v^- = 0$ we then obtain
\begin{equation}
p^- + (r_1^- -r_2^- + r_3^-) q^- = 0\,,
\label{e:pBC}
\end{equation}
which determines $p^-$.
Then, taking the limit of~\eqref{e:kpwhitham} as $x\to-\infty$ yields
\bse
\label{e:BCs}
\begin{align}
\label{e:BC1}
&\partialderiv{r_j^-\!}{t} + 2 \lambda q^-  \ddy{r_j^-} + \lambda \nu_j^- \ddy{q^-} + \lambda \ddy{p^-}
  = 0\,, \qquad j=1,2,3,
\\
\label{e:BC2}
&\partialderiv{q^-\!}{t} + 2 \lambda q^- \ddy{q^-} + \nu_{4.1}^- \ddy{r_1^-} + \nu_{4.3}^- \ddy{r_3^-} = 0\,,  
\\
\noalign{\noindent which determine the time evolution of $r_1^-,\dots,r_3^-$ and $q^-$, together with}
\label{e:BC3}
&(1-\alpha^-)\ddy{r_1^-} - \alpha^- \ddy{r_3^-} = 0\,,
\end{align}
\ese
which would seem to impose a constraint on the set of admissible BCs.
We next show, however, that when $m^-=0$ or $m^-=1$, \eqref{e:BCs} is a self-consistent system.

Recall that, when $r_1^-=r_2^-$, one has $m^-=0$, and the coefficients of the KP-Whitham system~\eqref{e:kpwhitham} assume a particularly simple form.
In particular, $\alpha^-=1$,
\eqref{e:BC3} and~\eqref{e:BC1} with $j=3$ yield respectively $\partial r_3^-/\partial y = 0$ 
and $\partial r_3^-/\partial t = 0$ (as it should be since $u^- = r_3^-$).
Moreover, the PDEs obtained from~\eqref{e:BC1} with $j=1,2$ coincide (as it should be since $r_1^- = r_2^-$).
Finally, \eqref{e:BC1} with $j=1$ and~\eqref{e:BC2} yield the following system of 2 (1+1)-dimensional ODEs for
$r^- = r_1^-$ and $q^-$:
\begin{equation}
\label{e:kpwhithamleft3}
\partialderiv{r^-\!}t + 2 \lambda q^- \partialderiv{r^-\!}y = 0\,,
\qquad
\partialderiv{q^-\!}t + 2 \lambda q^- \partialderiv{q^-\!}y = 0\,,
\end{equation}
which determine completely the time evolution of $r_1^-$ and $q^-$.

Similarly, when $r_2^-=r_3^-= r^-$, one has $m^- = 1$ and $\alpha^-=0$. 
Hence~\eqref{e:BC3} and~\eqref{e:BC1} with $j=1$ yield respectively $\partial r_1^-/\partial y = 0$ 
and $\partial r_1^-/\partial t = 0$ (as it should be since $u^- = r_1^-$).
Moreover, 
the PDEs obtained from~\eqref{e:BC1} with $j=2,3$ coincide (as it should be since $r_2^- = r_3^-$).
Finally, \eqref{e:BC1} with $j=3$ and~\eqref{e:BC2} yield the following system of 2 (1+1)-dimensional ODEs for
$r^- = r_3^-$ and $q^-$:
\bse
\label{e:kpwhithamright3}
\begin{align}
&\partialderiv{r^-\!}{t} + 2 \lambda q^- \partialderiv{r^-\!}{y} + \frac{4}{3}\lambda (r^- - u^-)\partialderiv{q^-}{y}= 0\,,
\\
&\partialderiv{q^-}{t} + 2 \lambda q^- \partialderiv{q^-}{y} + 4 \partialderiv{r^-}{y} = 0\,,
\end{align}
\ese

{Similar considerations apply for the BCs as $x\to\infty$.}
That is, \eqref{e:kpwhithamleft3} or~\eqref{e:kpwhithamright3} 
(as appropriate in the specific case) 
hold as $x\to\infty$ when $r^-$ and $q^-$ are replaced by $r^+$ and $q^+$.
Note that the Hopf equation for $q^-$ in~\eqref{e:kpwhithamleft3} has the same form as that for $q$ in the cKdV reduction [cf.~\eqref{e:qHopf}]. 
For the KPII equation, nondecreasing initial-boundary conditions of the form $q^\pm(y,0) = c_oy^{2n+1}$
(with $c_o$ a positive constant and $n$ a positive integer and $n=1$ corresponding to the cKdV reduction), or suitable combinations thereof, 
will not develop a shock singularity at $t>0$.

\paragraph{Riemann problems for the KP-Whitham system.}
We now turn out attention more specifically to the (2+1)-dimensional generalization of the Riemann problem for the KdV equation.
More precisely, we consider solutions of the KP-Whitham system~\eqref{e:kpwhitham} with initial conditions corresponding to a single front.
As in the one-dimensional case, one typically needs to solve the Whitham system with regularized initial conditions for the Riemann-type variables $r_1,\,r_2,\,r_3$ and $q$ 
and {then} compare the numerical results with the direct numerical simulations of the KP equation
{to verify that the KP-Whitham system yields a faithful approximation of the dynamics.}
For brevity, in this paper we limit ourselves to introducing and discussing the methods that can be used to solve the problem. 
The numerical simulations and the comparisons between the results of the KP-Whitham systems and direct numerical simulations of the KPI/KPII equations will be discussed elsewhere.

Consider initial conditions in the form of a generic single front specified by $x + c(y) = 0$ where $c(y)$ is an arbitrary function of $y$. 
Accordingly, 
we consider a step-like initial datum for $u$ as 
\vspace*{-1ex}
\begin{align}
\label{e:frontic}
u(x,y,0) = \begin{cases}
1, \quad x + c(y)<0\,, \\
0, \quad x + c(y) \geq 0\,,
\end{cases}
\end{align}
{where the values 1 and 0 can be selected without loss of generality thanks to the Galilean invariance of the KP equation and the KP-Whitham system.}
If $c(y)$ is constant or linear in $y$, the setting obviously reduces to Riemann problem for the KdV equation.
Also, if $c(y)$ is a quadratic function of $y$ the setting reduces to the Riemann problem for the cylindrical KdV equation.

Similarly to the case of the KdV equation 
{and the cKdV equation \cite{ckdv}, 
it is convenient to regularize the jump and choose}
the corresponding initial conditions for the Riemann variables $r_1,\,r_2$ and $r_3$ to be 
\begin{align}
\label{e:r-ic}
r_1(x,y,0) = 0, \quad r_2(x,y,0) = R_2(x+c(y)), \quad r_3(x,y,0) = 1\,.
\end{align}
where {the IC for $r_2$ ``regularizes'' the jump by interpolating smoothly between the values 0 and~1; e.g.,}
${R_2(\eta)} = \frac12 \big(1+\tanh\big[{\eta}/\delta\big]\big)$
where $\delta$ is a small parameter.
{To determine the corresponding IC for the fourth variable,}
note that the ICs~\eqref{e:r-ic} imply that the constraint~\eqref{e:constraint} is satisfied at $t=0$.
Then, from~\eqref{e:kdef} and~\eqref{e:r-ic}, we have in this case 
\begin{align*}
k = 1/(2 \K(r_2))\,,
\end{align*}
and it is easy to check that $k_y(x,y,0) = c'(y) k_x(x,y,0)$.
Therefore, substituting in~\eqref{e:q-ic}, the IC for $q$ simply reduces to
\begin{align}
q(x,y,0) = c'(y)\,.
\end{align}
Again, if $c(y)$ constant or linear in $y$ the IC for $q$ is trivial, whereas if $c(y)$ is a quadratic function of $y$ one reduces to the ICs of the Riemann problem for the cylindrical KdV equation.
{The IC for $p$ is chosen as described earlier,} {namely via~\eqref{e:pintegrated} at $t=0$ and~\eqref{e:pBC}.}

{Based on the above discussion, one expects that 
simple ICs that lead to (2+1)-dimensional DSWs for the KPII equation might take the form $c(y) = c_o y^{2n}$ 
or suitable combinations thereof, 
with $n$ a positive integer and $c_o$ a positive constant. 
The cKdV reduction, obtained with $n=1$, is the simplest type of such ICs, and does indeed generate (2+1)-dimensional DSWs~\cite{ckdv}.}

\section{Stability analysis of the periodic solutions of the KP equation}
\label{s:stability}

Here we show how the KP-Whitham system~\eqref{e:kpwhitham} can also be used to investigate the stability properties of the genus-1 (i.e., cnoidal, or traveling-wave) solutions of the KP equation.

Recall that, for an exact cnoidal wave solution of the KP equation, the Riemann invariants (as well as {$p$ and} $q$) are constants in time as well as independent of $x$ and $y$.
To investigate the stability of the cnoidal wave, we consider a small initial perturbation of the Riemann invariants, {$p$} and $q$ and use
the KP-Whitham system~\eqref{e:kpwhitham} to study the evolution of such a perturbation.
That is, we look for solutions of~\eqref{e:kpwhitham} in the form
\begin{align}
\label{e:perturb}
r_1 = \~r_1 + r_1'\,, \quad r_2 = \~r_2 +  r_2'\,, \quad r_3 = \~r_3 +  r_3'\,, \quad q = q'\,, \quad {p = p'\,,}
\end{align}
where $\~r_1,\~r_2,\~r_3$ are arbitrary constants, satisfying $\~r_1 \leq \~r_2 \leq ,\~r_3$
and where we have set {$\~p=0$} and $\~q=0$ without loss of generality using the invariances of the KP equation.
We then seek a perturbation expansion with $|r_j'(x,y,t)| \ll 1$ for $j = 1,2,3$, {$|p'(x,y,t)| \ll 1$} and $|q'(x,y,t)| \ll 1$.

Substituting~\eqref{e:perturb} into the KP-Whitham system~\eqref{e:kpwhitham} and dropping higher-order terms, we have
\vspace*{-1ex}
\bse
\label{e:linearterm}
\begin{align}
&\partialderiv{r_j'}{t} + \~V_j \,\partialderiv{r_j'}{x} + \lambda \~\nu_j \partialderiv{q'}{y} 
  { + \lambda \partialderiv{p'}{y}} = 0\,,\qquad j=1,2,3,\\
&\partialderiv{q'}{t} + \~V_2\,\partialderiv{q'}{x}
+ \~\nu_{4.1} \partialderiv{r_1'}{y} + \~\nu_{4.3} \partialderiv{r_3'}{y} = 0\,,  \\
&{ \partialderiv{p'}{x} - (1-\~\alpha) \partialderiv{r_1'}{y} - \~\alpha \partialderiv{r_3'}{y} + \~\nu_5     \partialderiv{q'}{x} = 0\,,}
\end{align}
\ese
where $\~V_1,\dots,\~V_3$, $\~\nu_1,\dots,\~\nu_3,\~\nu_{4.1}\,,\~\nu_{4.3}\,,$ {$\~\nu_5$ and $\~\alpha$} denote the unperturbed values of all the corresponding coefficients, as defined in~\eqref{e:Vdef} and~\eqref{e:nudef}.
(I.e., the value of those coefficients for the unperturbed solution).
Next we look for plane wave solution of the above system of linear PDEs in the form 
\begin{align}
\label{e:planewave}
r_j'(x,y,t) = R_j \,\e^{i(K x + L y - W t)}\,,~ j = 1,2,3, \qquad 
{\big( q'(x,y,t) , p'(x,y,t) \big) = (Q,P)\,\e^{i(K x + L y - W t)}\,.}
\end{align}
Substituting~\eqref{e:planewave} into~\eqref{e:linearterm} yields
{the homogeneous linear algebraic system}
\bse
\label{e:linearterm2}
\begin{align}
&(W - K \~V_j) R_j = \lambda L \~\nu_j Q { + \lambda L P}\,,\qquad j=1,2,3, \\
&(W - K \~V_2) Q = L \~\nu_{4.1} R_1 + L \~\nu_{4.3} R_3\,, \\
&{K P = L (1-\~\alpha) R_1 + L \~\alpha R_3 - K \~\nu_5 Q\,.}
\end{align}
\ese
{Non-trivial solutions for the Fourier amplitudes $(R_1,R_2,R_3,Q,P)$ exist when the determinant of the corresponding coefficient matrix vanishes,}
which in turn yields the linearized dispersion relation 
{
\begin{align}
\label{e:dispersionrelation}
f_4(K,L,W) = 0\,,
\end{align}
where $f_4(K,L,W)$ is a cubic polynomial in $W$ and quartic in $K$ and $L$.}
The cnoidal wave solution of KP corresponding to $\~r_1,\~r_2,\~r_3$ will be linearly stable 
if all solutions of~\eqref{e:dispersionrelation} are real (because in this case perturbations remain bounded),
whereas if~\eqref{e:dispersionrelation} admits solutions with non-zero imaginary part, 
some perturbations will grow exponentially, implying that the cnoidal wave is unstable.

Studying analytically the solutions of~\eqref{e:dispersionrelation} is nontrivial.
However, we can obtain a much more tractable situation by taking $K = 0$, 
i.e., by considering perturbations that are independent of~$x$.
Physically, taking $K=0$ corresponds to considering slowly varying perturbations of the cnoidal wave in the transverse direction.
Then~\eqref{e:dispersionrelation} simplifies to
\vspace*{-0.6ex}
\begin{gather}
\label{e:dispersionrelation2}
(W/L)^2 = \lambda f(r_1,r_2,r_3)\,,
\\
\noalign{\noindent where}
{f(r_1,r_2,r_3) = (\~\nu_3 - \~\nu_1)(\~\nu_{4,3} (1-\~\alpha) - \~\nu_{4.1} \~\alpha)\,.}
\label{e:Ndef}
\end{gather}
The necessary criterion for the linear stability of the cnoidal wave is now apparent:
the cnoidal wave solution of KP corresponding to {the constant unperturbed values} $\~r_1,\~r_2,\~r_3$ can be linearly stable 
if the RHS of~\eqref{e:dispersionrelation2} is non-negative.
Conversely, if the RHS of~\eqref{e:dispersionrelation2} is negative, $W$ is purely imaginary, implying that the unperturbed solution is unstable.
Note that, for this particular case, the stability properties of the solutions of KPI ($\lambda=-1$) and KPII ($\lambda=1$) are necessarily opposite:
if solutions of one are stable, solutions of the other are unstable and vice versa.

We can further simplify the problem by considering cnoidal waves with
$\~r_1 = 0$ and $\~r_3 = 1$.
{Note that we can do so without loss of generality thanks to the invariance of the KP-Whitham system under scaling and 
Galilean transformations.}
In this case the elliptic parameter is simply $m = \~r_2$, and $f(r_1,r_2,r_3) = f(m)$, with
\begin{equation}
f(m) = \frac{4(3\E^2-2(2-m)\E\K + (1-m)\K^2)^2}{3\E\K(\K-\E)(\E-(1-m)\K)}\,.
\label{e:growthrate}
\end{equation}
{It is straightforward to see that $f(0)=0$ and $f(m)>0$ for all $0< m\le 1$.}
As a result, for the KPI equation ($\lambda = -1$), $W$ is purely imaginary, 
and therefore all of its cnoidal waves are linearly unstable. 
In contrast, for the KPII equation ($\lambda = 1$), $W$ is real-valued, and therefore all of its cnoidal waves are linearly stable.
(Note however that the stability result for KPII is limited to the framework of this analysis,
namely $K=0$.
To determine the full linear stability properties for KPII one would have to prove that $W$ is real for all values of $K$.)
As $m\to1$, cnoidal waves become line solitons, and we recover the well-known result that the line soliton solutions of KPI are unstable to slowly varying transverse perturbations~\cite{AS1981}. 
The above discussion, however, generalizes this instability result to cnoidal wave solutions with arbitrary $m$.

\begin{figure}[t!]
\centering
\includegraphics[width=0.455\textwidth]{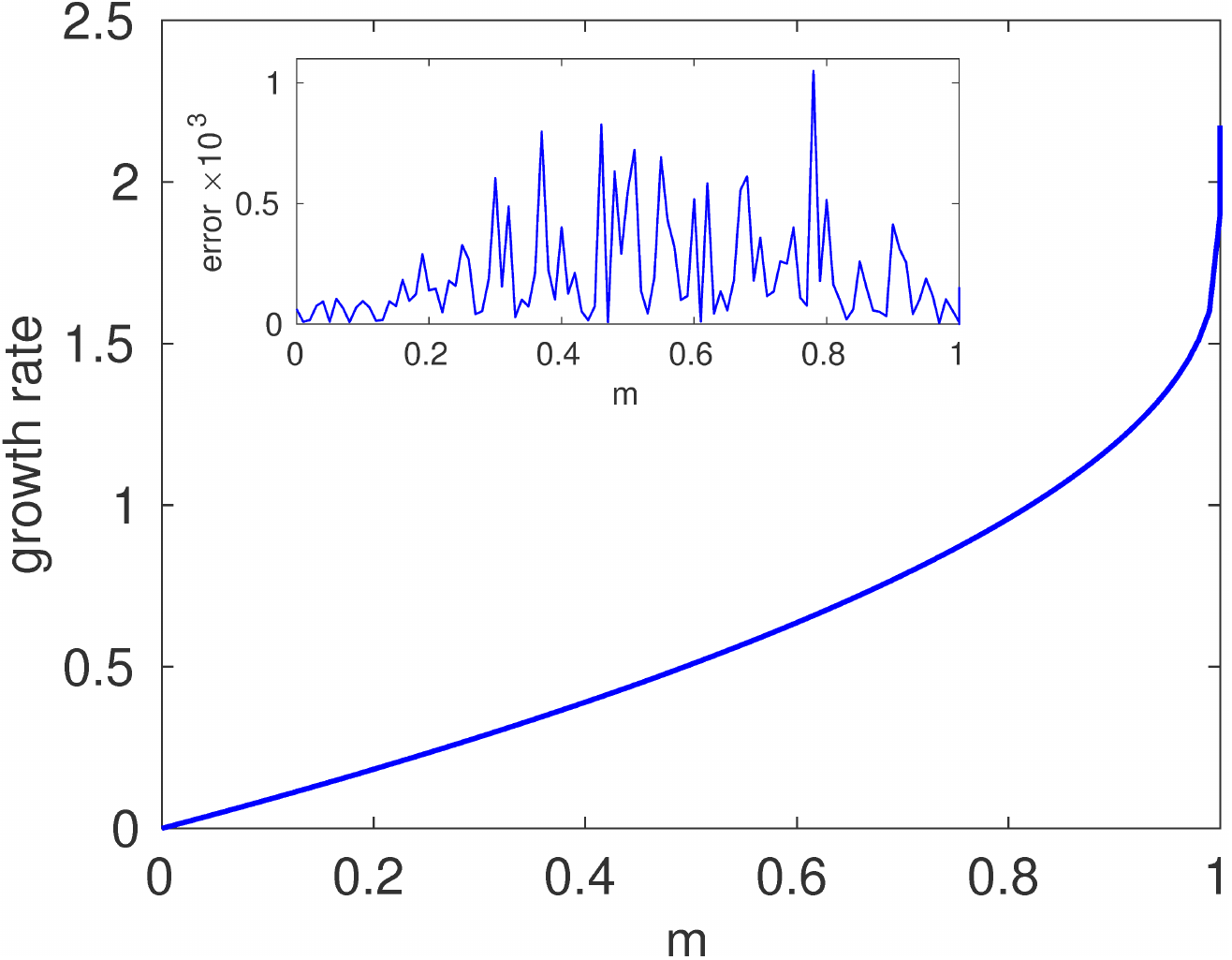}\quad%
\includegraphics[width=0.505\textwidth]{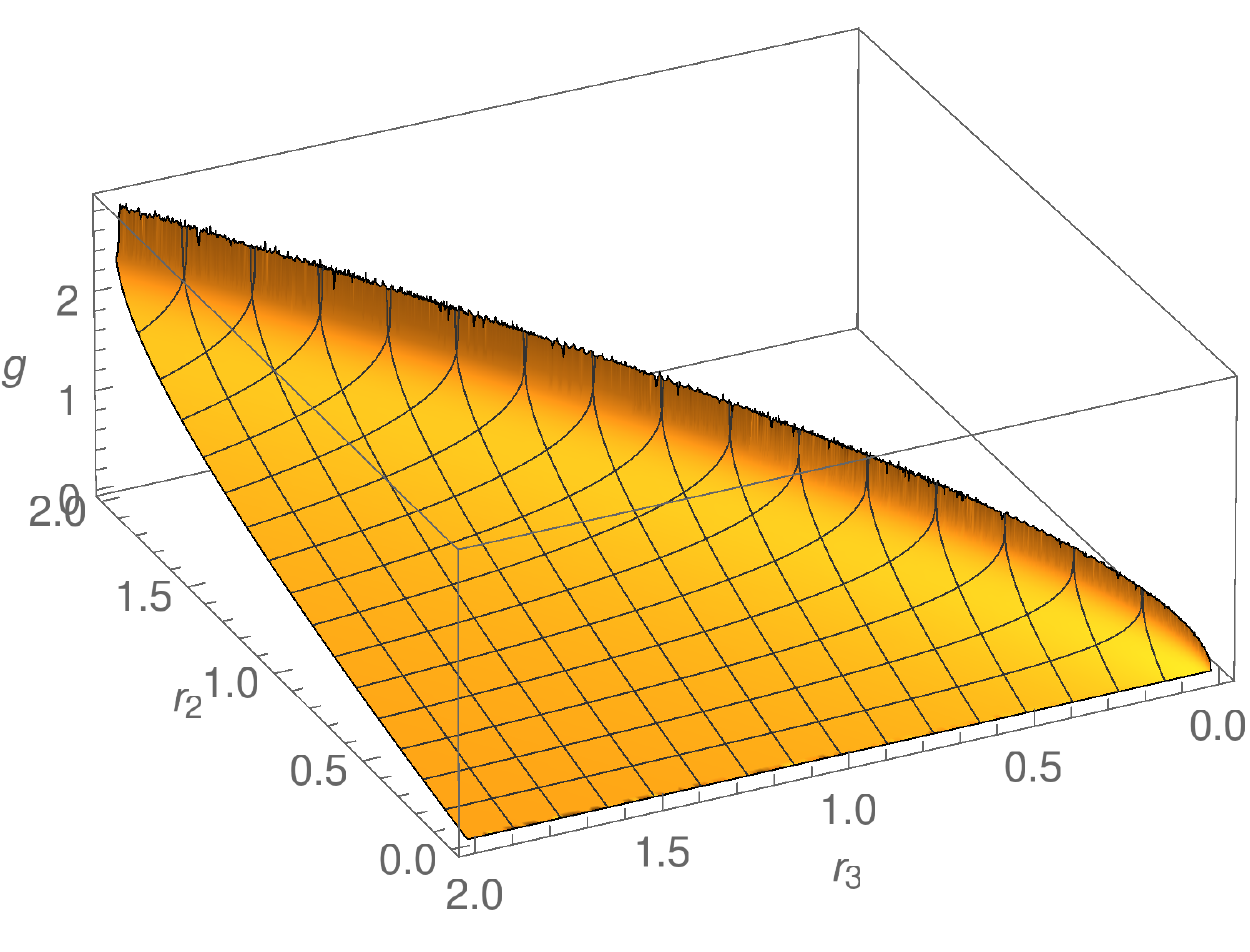}
\caption{
Left: The value of {$g(m)=\sqrt{f(m)}$} from~\eqref{e:growthrate} as a function of $m = r_2$ when $r_1=0$ and $r_3=1$.
{The inset shows the difference between the growth rate as determined from Whitham theory and from direct numerical simulations 
of the linearized KP equation around a cnoidal wave solution.}
Right: The {growth rate $g$} from~\eqref{e:Ndef} as a function of $r_2$ and $r_3$ when $r_1=0$.}
\label{f:N}
\vglue-2\smallskipamount
\end{figure}

{Figure~\ref{f:N}(left) shows the value of the growth rate $g(m) = \sqrt{f(m)}$ as a function of $m$ for $0\leq m\leq 1$.}
Interestingly, Fig.~\ref{f:N} also shows that $g(m)$ is a monotonically {increasing} function of $m$ between 
{$g(0) = 0$ and $g(1) = 4/\sqrt3$.}
Note in particular that the value $g(1) = 4/\sqrt3$ coincides with the growth rate of the unstable perturbations that is obtained
from a direct linearization of the KPI equation around its line soliton solutions~\cite{AS1981}.
The fact that $g(m)$ is monotonically {increasing} with $m$ also indicates that the solitonic sector for KPI ($m$ close to 1) 
is {more} unstable than the cnoidal wave sector, which in turn is {more} unstable than the linear sector ($m$ close to 0).
Indeed, the fact that $g(0)=0$ implies that the constant background of the KPI is linearly stable, consistently with the results
of a direct linear stability analysis.
Interestingly, it is also possible to analytically compute the slope of the curve $g(m)$ at $m=0$, to obtain $g'(0) = 2/\sqrt3$.

{It should be noted that partial results regarding the stability/instability properties of the cnoidal-wave solutions of KPI/KPII 
had already been obtained in a few existing studies \cite{hakkaev,haragus,kuznetsov}.} 
{The analytical {growth rate} estimate~\eqref{e:growthrate}, however,} is novel to the best of our knowledge. 

As a slightly more general case, we can look at $f(r_1,r_2,r_3)$ as a function of $r_2$ and $r_3$ when $r_1=0$.
Figure~\ref{f:N}(right) shows the value of {the growth rate as a function} of $r_2$ and $r_3$ 
(with $r_2 \leq r_3$ as required for consistency with the KP-Whitham system). 
From Fig.~\ref{f:N}(right), one can see that the value of $g(r_2,r_3) = \sqrt{f(r_2,r_3)}$ is always positive. 
Therefore, the conclusions of the previous paragraphs hold true in this more general scenario.

To check the results from Whitham theory, we also computed the {growth rates} for the KPI equation 
by direct numerical evaluation of the spectrum of the linearized KPI equation around its cnoidal wave solutions 
using Floquet-Fourier-Hill's methods similarly to \cite{deconinckkutz}
(see Appendix~\ref{s:directstability} for details). 
The difference between the growth rates obtained from the numerical simulations and those predicted from Whitham theory is shown in the inset of Fig.~\ref{f:N}\,(left). 
It is evident from the figure that the agreement is excellent,
which provides a strong indication of the validity of the perturbation expansion presented in Section~\ref{s:derivation} 
and confirms the usefulness of the KP-Whitham system itself.

It is important to note that ignoring the last PDE in the Whitham system and setting $p={}$const would yield incorrect stability results.
Specifically, one would still obtain that the cnoidal wave solutions of KPI are unstable and those of KPII are stable,
but the resulting growth rate $g(m)$ for KPI would be a decreasing function of $m$ instead of an increasing one,
and one would also get $g(0)\ne0$ implying that a constant background of the KPI is unstable, 
contrary to the results of a direct linearization.

We can also consider perturbations of the similarity solution of the KdV-Whitham system found in \cite{GP1974}, 
which describe a DSW for the KdV equation.
In this case one has $\~r_1=0$, $\~r_2=\~r_2(x/t)$ and $\~r_3=1$.
Therefore, in the KP-Whitham system~\eqref{e:kpwhitham} we look for solutions in the form
\begin{align}
\label{e:perturbdsw}
r_1 =  r_1'\,, \quad r_2 = \~r_2 +  r_2'\,, \quad r_3 =1+  r_3'\,, \quad q = q'\,, \quad {p = p'\,} 
\end{align}
with $\~r_2 = \~r_2(\xi)$ and $\xi=x/t$.
Substituting into~\eqref{e:kpwhitham} and linearizing the resulting equations,
we find the following (2+1)-dimensional system of PDEs in the independent variables $\xi,y$ and $t$:
\bse
\label{e:linearterm3}
\begin{align}
&\partialderiv{r_j'}{t} -\frac{\xi}{t}\frac{\partial r_j'}{\partial \xi} + \frac{\~V_j }{t}\,\frac{\partial r_j'}{\partial \xi} 
  + \lambda \~\nu_j \partialderiv{q'}{y}
  { + \lambda \partialderiv{p'}{y}} = 0\,,\qquad j=1,2,3,\\
&\partialderiv{q'}{t}  -\frac{\xi}{t}\frac{\partial q'}{\partial \xi} + \frac{\~V_2}{t}\,\frac{\partial q'}{\partial \xi}
+ \~\nu_{4.1} \partialderiv{r_1'}{y} + \~\nu_{4.3} \partialderiv{r_3'}{y} = 0\,, \\
&{\frac{1}{t} \frac{\partial p'}{\partial \xi} - {(1-\~\alpha)} \partialderiv{r_1'}{y} - {\~\alpha} \partialderiv{r_3'}{y} + \frac{\~\nu_5}{t} \frac{\partial q'}{\partial \xi} = 0\,,}
\end{align}
\ese
where all unperturbed values are now functions of $\xi$.
For all finite values of~$\xi$, the terms proportional to $1/t$ decay as $t\to\infty$, and we recover the same linearized system as above, 
namely~\eqref{e:linearterm} in the special case $K=0$, but with $\xi$ as a parameter. 
Therefore, the same results apply. This indicates that the DSW itself is unstable.
This result should not be surprising in light of the results of this section
(namely, the fact that each ``elliptic function component'' of the DSW is unstable).

\section{Concluding remarks}
\label{s:remarks}

The results of this work open up a number of interesting questions, both from a mathematical and from a physical point of view.

1. From a theoretical point of view, a natural question is whether the KP-Whitham system~\eqref{e:kpwhitham} is completely integrable.
Note that~\eqref{e:kpwhitham} is an {asymptotic} reduction of the KP equation, which is itself an integrable system,
Hence one would suspect that the KP-Whitham system~\eqref{e:kpwhitham} is integrable. 
For a (2+1)-dimensional system of PDEs of hydrodynamic type, 
the integrability condition involves the Ferapontov-Khusnutdinova test~\cite{ferapontov2006},
which identifies the vanishing of the Haantjes tensor as a necessary condition for integrability.
Interestingly, the system~\eqref{e:kpwhitham} does not pass this test.
A more general test for integrability exists, 
involving the direct search for existence of hydrodynamic-type reductions with an 
arbitrary number of components \cite{ferapontov2004}.
Such a calculation is outside the scope of this work, however.

{2.} If the KP-system {is integrable, an important} question would then be whether one could 
formulate a method to solve the initial value problem (IVP) possibly using the novel generalization of the inverse scattering transform for vector fields that was recently developed by Manakov and Santini~\cite{manakovsantini1,manakovsantini2,manakovsantini3}
to solve the IVP for dispersionless systems.

{3.} Another interesting question related to the integrability of the KP equation is the derivation of 
KP-Whitham equations of higher genus.
Note that a formal modulation theory for the KP equation was presented in \cite{krichever} 
using the Riemann surface machinery for the finite-genus solutions of the KP equation.
In this formalism, the Whitham modulation equations of arbitrary genus are obtained by averaging the conservation laws
of the integrable PDE over the fast variables.
In principle, these methods should allow one to recover the genus-1 KP-Whitham system~\eqref{e:kpwhitham} 
as well as to obtain all of its higher-genus generalizations.

{4.} Yet another question is whether there exist further, more general exact reductions of the system~\eqref{e:kpwhitham} 
other than those to the KdV, slanted KdV and cKdV equations.
Note that, of the three reductions discussed in section~\ref{s:properties}\ref{s:reductions}, the first two are such that $Dq/Dy$ vanishes identically,
whereas the third one is such that $Dq/Dy$ is a function of $t$.
The question is then whether there are more general situations that yield similar conditions.
This issue might also be related to integrability, 
since the definition of integrability for a hydrodynamical system of PDEs according to \cite{ferapontov2004} 
is the existence infinitely many suitable reductions.

{5.} On the other hand, we emphasize that 
none of the results of this work depend on the fact that the KP equation itself is integrable.
Therefore, the methods used in this work are be applicable to other (2+1)-dimensional PDEs. 
Indeed, we have also used the same methods to formulate the Whitham modulation equations for the 
(2+1)-dimensional generalization of the Benjamin-Ono equation, which is not integrable.
Those results will be reported as a separate publication.

6. In fact, many important questions about the KP-Whitham system~\eqref{e:kpwhitham} are independent of whether the system is integrable.
From an analytical point of view, one such question is whether there are any rigorous conditions for the global existence of solutions 
of the KP-Whitham system~\eqref{e:kpwhitham} which generalize those available for the KdV-Whitham system 
(namely, the result that if the ICs for the Riemann invariants $r_1,r_2,r_3$ are non-decreasing, the KdV-Whitham system admits a global solution, as a consequence of the the sorting property of the velocities $V_1,V_2,V_3$.)

7. From a practical point of view, 
{an} opportunity of future study will be to {perform} careful numerical simulations of the KP-Whitham system~\eqref{e:kpwhitham} 
with a variety of ICs (especially ones that cannot be reduced to one-dimensional cases) 
and {carry out a detailed} comparison with the original PDE (i.e., the KP equation).

{8.} {A related question is whether one can use} the KP-Whitham system to {regularize the} singularity of the 
genus-0 system (i.e., the un-regularized, dKP equation), {and thereby} compare the development of the gradient catastrophe 
in the dispersionless system~\cite{grava} to {the behavior of solutions of the regularized system~\eqref{e:kpwhitham} and of the KP equation itself.}
(For example, it was shown in~\cite{breaking1,breaking2,breaking3} that the initial singularity for the dKP system
arises at a single point.  It is an open question whether the same result carries over to the regularized system.)

{9.} The instability of the genus-1 solutions of the KPI equation raises the question of whether the corresponding genus-1 {KPI-Whitham} 
system can ever admit nontrivial regular solutions, or whether instead the initial shock must be regularized by a more general, yet to be derived, 
{higher-genus} KP-Whitham system.
(Note in this respect that the instability of the line solitons of KPI results in the formation of a periodic array of lumps \cite{stepanyants}, and such a structure cannot be captured as limits of genus-1 solutions, which are all one-dimensional objects.)

{10.}
Finally, it should be noted that a simplified derivation of the Whitham system~\eqref{e:kpwhitham} can be given, and will be reported separately.
An equivalent system can also be obtained when (1.4b) is replaced with the slightly simpler PDE
\begin{equation}
\frac{\partial q}{\partial t} + (V + \lambda q^2)\,\frac{\partial q}{\partial x} + \frac{D}{Dy}\big( V + \lambda q^2 \big) = 0\,, 
\end{equation}
with $V$ defined by~(2.18) and given by~(2.25) as before.
(Note also that, even though taking $p$ to be constant in (1.4a) with (1.4b) replaced by (5.1) would yield a formally different $4\times4$ reduction from the one discussed here,
the stability results of both $4\times4$ reductions are identical.  As mentioned before, the predictions
of the $4\times4$ reduction are inconsistent with those of the full $5\times5$ system and with the results from direct numerical simulations.)

It is hoped that the results of this work and the above discussion will stimulate further work on these problems.

\section*{Acknowledgments}

We thank Ali Demirci and Justin Cole for detailed and helpful interactions and Eugeny Ferapontov, Mark Hoefer and Antonio Moro
for many insightful discussions related to this work.
This work was partially supported by the National Science Foundation under grant numbers  DMS-1310200 and DMS-1614623.

\section*{Appendices}
\setcounter{section}0
\def\thesubsection{\Alph{subsection}}
\renewcommand{\theequation}{\Alph{subsection}.\arabic{equation}}
\addcontentsline{toc}{section}{Appendix}

\subsection{The KdV-Whitham system}
\label{s:riemannkdv}

Here we briefly review the derivation of the Riemann invariant variables of the Whitham system for the KdV equation.
We start from the three modulation equations, which can be obtained from the the first three of~\eqref{e:meqns} in the main text
by taking $\lambda=0$ and removing the dependence with respect to $y$.
The result are three quasi-linear (1+1)-dimensional PDEs for the independent variables $V$, $\beta=b/m$ and $m$. 
Next, we show that one can introduce the Riemann invariant variables to diagonalize the above system. 
Introducing the notation 
\begin{align}
\label{e:kdv-notation}
\mathbf{w}(x,t) = (w_1\,, \, w_2\,, \, w_3)^T := (V\,,\,b/m\,,\,m )^T,
\end{align}
for brevity, we rewrite the system of modulation equations as a vector system of equations
\begin{align}
\label{e:kdv-prematrix}
R\,{\mathbf{w}}_t + S\,{\mathbf{w}}_x = 0\,,
\end{align}
or equivalently,
\vspace*{-1ex}
\begin{align}
\label{e:kdv-matrixeq}
{\mathbf{w}}_t + A\,{\mathbf{w}}_x = 0\,,
\end{align}
where $A = R^{-1}S$.
(Explicitly, the entries of $A$ are obtained from the first three rows and columns of the corresponding matrix for the 
KP equation by setting $\lambda = q = 0$; cf.\ Appendix~\ref{s:matrix}.)
To diagonalize the system~\eqref{e:kdv-matrixeq}, one must diagonalize the matrix $A$. 
Straightforward calculations show that 
\bse
\label{e:kdv-diagonal}
\begin{align}
\label{e:kdv-diagA}
A = P^{-1} D P \,,
\end{align}
where
\begin{align}
\label{e:kdv-diagP}
P = \frac{1}{3 w_2}
\begin{pmatrix}
w_3 -1 & 1 - w_3^2 & (1 - w_3) w_2 \\
1 & 2 w_3 - 1 & 2 w_2 \\
-w_3 & w_3 (w_3 - 2) & w_3 w_2
\end{pmatrix}\,
\end{align}
\ese
and $D = \mathrm{diag}(V_1,V_2,V_3)$, with $V_1,V_2,V_3$ given by~\eqref{e:Vdef} in the main text.
Using~\eqref{e:kdv-diagA}, we can then write~\eqref{e:kdv-matrixeq} in the main text as
\begin{align}
\label{e:kdv-matrixeq2}
P\,{\mathbf{w}}_t + D P\,{\mathbf{w}}_x = 0\,.
\end{align}
The key to find the Riemann invariants is to find dependent variables $r_1, r_2$ and $r_3$ such that
\begin{align}
\label{e:kdvcondition}
P \,{\mathbf{w}}_x = {\mathbf{r}}_x\,,\qquad
P \,{\mathbf{w}}_t = {\mathbf{r}}_t\,,
\end{align}
with $\mathbf{r} = (r_1,r_2,r_3)^T$.
After canceling out common factors in each row of the LHS of both parts of~\eqref{e:kdvcondition}, the three rows in the LHS of~\eqref{e:kdvcondition} 
can be written as
\begin{align*}
w_{1,x} - (1+w_3) w_{2,x} - w_2\,w_{3,x} = (w_1 - w_2\ - w_2 w_3)_x\,, \\
w_{1,x} + (2w_3 - 1) w_{2,x} + 2 w_2\,w_{3,x} = (w_1 - w_2\ + 2 w_2 w_3)_x\,, \\
w_{1,x} + (2-w_3) w_{2,x} - w_2\,w_{3,x} = (w_1 + 2 w_2\ - w_2 w_3)_x\,,
\end{align*}
plus identical expressions for the temporal derivatives.
Finally, taking
\begin{align}
r_1 = (w_1 - w_2\ - w_2 w_3)/6\,, \quad r_2 = (w_1 - w_2\ + 2 w_2 w_3)/6\,, \quad r_3 = (w_1 + 2 w_2\ - w_2 w_3)/6\,,
\end{align}
and solving for $w_1,w_2,w_3$ we obtain~\eqref{e:riemannvariables}.
That is, the change of variables~\eqref{e:riemannvariables} transforms
the system of equations~\eqref{e:kdv-prematrix} into the diagonal system~\eqref{e:whithamkdv}.


\subsection{Stability analysis of periodic solutions via direct numerical simulations}
\label{s:directstability}

In this section we briefly discuss the calculations of the stability properties of the cnoidal wave solutions for the KP equation
by direct numerical simulations.

Let $u(x,y,t) = u_0(\xi)$ be a traveling wave (a.k.a. elliptic, periodic, cnoidal-wave or genus-1) solution of the KP equation~\eqref{e:kp} with $\xi = x-c t$.
(Note that without loss of generality we can always align any traveling wave solution along the $x$-axis thanks to the pseudo-rotation invariance of the KP equation.)
Next, consider a perturbed solution of the KP equation in the form $u(x,y,t) = u_0(\xi) + v(\xi,y,t)$ with $|v(\xi,y,t)| \ll 1$. 
Substituting into the KP equation~\eqref{e:kp} and dropping higher-order terms, we have
\begin{align*}
\Big(v_t - c v_{\xi} + 6 (u_0 v)_{\xi} + \epsilon^2 v_{\xi\xi\xi}\Big)_{\xi} + \lambda v_{yy} = 0\,.
\end{align*}
Using the Galilean invariance of the KP equation, we can always perform the transformation $u_0 \mapsto c/6 + \~u_0$, which yields
\begin{align}
\label{e:perturbeqn}
\Big(v_t + 6 (\~u_0 v)_{\xi} + \epsilon^2 v_{\xi\xi\xi}\Big)_{\xi} + \lambda v_{yy} = 0\,.
\end{align}
Now we look for plane wave solution of the above equation in the form
\begin{align}
\label{e:planewavesoln}
v(\xi,y,t) = w(\xi) \e^{i \zeta y + \mu t}\,.
\end{align}
Substituting~\eqref{e:planewavesoln} into~\eqref{e:perturbeqn} yields
\begin{align*}
\Big(\mu w + 6 (\~u_0 w)_{\xi} + \epsilon^2 w_{\xi\xi\xi}\Big)_{\xi} - \lambda \zeta^2 w = 0\,,
\end{align*}
or equivalently,
\begin{align}
\label{e:eigenprob}
\mu w + 6 (\~u_0 w)_{\xi} + \epsilon^2 w_{\xi\xi\xi} - \lambda \zeta^2 \partial^{-1}_{\xi} w = 0\,,
\end{align}
where $\partial^{-1}_{\xi} w = \mathcal{F}^{-1} \Big[(1/i k)\,\mathcal{F}[w]\Big]$ with $\mathcal{F}$ denoting the Fourier transform.
Note that in order for~\eqref{e:eigenprob} to admit periodic solutions, one needs $\int^{\infty}_{-\infty} w(\xi) d \xi = 0$. 
That is,
$w$ should have zero mean (i.e., the Fourier transform of $w$ should vanish at the zero wave number).
Then we can finally write~\eqref{e:eigenprob} as an eigenvalue problem for a differential operator
\begin{align}
\label{e:eigenprob1}
- 6 (\~u_0 w)_{\xi} - \epsilon^2 w_{\xi\xi\xi} + \lambda \zeta^2 \partial^{-1}_{\xi} w = \mu w \,,
\end{align}
and solve it in the Fourier domain.
This operation is well-defined precisely because $w$ has no mean term.
In order to compare the results of this calculation with those obtained from the Whitham approach discussed in section~\ref{s:stability}, 
note that when $r_1 = q = 0$, $r_2 = m$ and $r_3 = 1$, the cnoidal wave solution~\eqref{e:ellipticsoln2} becomes
$
\~u_0 = 1 - m + 2 m \cn^2\big(x-(2+2 m)t,m\big)\,.
$
The eigenvalue problem was solved numerically with $0\le m \le 1$. 
The results, 
together with a numerical comparison with the Whitham approach, are given in Fig.~\ref{f:N}.


\subsection{Coefficient matrices for the un-regularized KP-Whitham system}
\label{s:matrix}

The entries of the coefficient matrix~$B$ of the original Whitham system~\eqref{e:matrixeqn} are given by:
\vspace*{-1ex}
\begin{align*}
&B_{11}= \lambda q~\frac{-2 b \big(\E(1+m)-\K(1+3m) \big) + (\E-\K) m V}{6 b \K m}\,,\\
&B_{12}= \lambda q~\frac{\big(\E-\K(1-m)\big) \big(2 b \big(-\K (1-m)^2 + \E (1+m) \big) - \big(\E - \K (1-m)\big) m V\big)}{6 b \K (\E-\K) (1-m) m}\,, \\
&B_{13}= \lambda q~\frac{\E \big(2 b \big( - 2 \K (1-m)+\E (1+m)\big)-\E m V\big)}{6 b (\K-\E) \K (1-m)}\,,\\
&B_{14}= \lambda~\frac{2b \big(\E(1+m)-\K \big)+(\K-\E) m V)}{6 (\K-\E) m}\,,\quad
{B_{15} = B_{25} = B_{35} =\lambda\,,}\\
&B_{21}= \lambda q~\frac{(\E-\K)(2 b \big(\K-\E (1-2 m)\big)+(\E-\K) m V)}{6 b \K \big(\E-\K (1-m)\big) m}\,, \\
&B_{22}= \lambda q~\frac{2 b \big(\K(1-5 m +4 m^2)-\E (1-2 m)\big)+\big(\E-\K (1-m)\big) m V}{6 b \K (-1+m) m}\,,\\
&B_{23}= \lambda q~\frac{\E \big(-2 b \big(\E(1- 2 m)- 2 \K (1-m)\big)+\E m V\big)}{6 b \K \big(\E-\K (1-m)\big) (1-m)}\,, \\
&B_{24}= \lambda~\frac{2b \big(\K (1-m)^2-\E (1-2 m))+\big(\E-\K (1-m)\big) m V}{6 \big(\E-\K (1-m)\big) m}\,,\\
&B_{31}= \lambda q~\frac{(\K-\E) \big(2 b \big(\K (2-3 m) - \E (2-m)\big)+(\K-\E) m V)}{6 b \E \K m}\,, & \\
&B_{32}= \lambda q~\frac{\big(\E-\K(1-m)\big) \big(2 b \big(\E (-2+m)-\K (-2+m+m^2)\big)-\big(\E-\K (1-m)\big) m V\big)}{6 b \E \K (1-m) m}\,, \\
&B_{33}= \lambda q~\frac{ 2 b \big(\E(2-m)+2\K(1-m)\big)-\E m V}{6 b \K (1-m)} \,,\\
&B_{34}= \lambda~\frac{-2b \big(\K(1-m)+\E (-2+m)\big)+\E m V}{6 \E m}\,, \\
&B_{41}= 2 + (V+\lambda q^2)\frac{\E-\K}{b \K} \,, \quad
B_{42}= 2 - (V+\lambda q^2)\frac{\E-\K (1-m)}{b \K (1-m)} \,, \\
&B_{43}= 2 + (V+\lambda q^2)\frac{\E m}{b \K (1-m)} \,, \quad
B_{44}=2 \lambda q\,, \quad B_{45}= B_{54}= B_{55}= 0\,, \\
&{B_{51}= -\frac{6 (\E-\K)^2}{\K^2 m}\,, \quad
B_{52}= -\frac{6 (\E- \K (1-m))^2}{\K^2 (-1+m) m}\,, \quad
B_{53}= - \frac{6 \E^2}{\K^2 (1-m)}\,.} 
\end{align*}
{Instead of listing all the entries of the coefficient matrix~$A$, we point out the important fact that,
even though both $A$ and $B$ are full matrices with complicated entries, the following combination of them takes on a particularly simple form:
\begin{align*}
A + q B = 
\begin{pmatrix}
 V_1 + \lambda q^2 & 0 & 0 & \gamma_1 & -\lambda q ~\frac{\K \big(2 b (1+m) + m V\big)}{6 b (\E-\K)}
\\
  0 & V_2 + \lambda q^2 & 0 & \gamma_2 & -\lambda q ~\frac{\K \big(2 b (1-2 m) + m V \big)}{6 b \big(\E -\K (1-m)\big)}
\\
 0 & 0 & V_3 + \lambda q^2 & \gamma_3 & -\lambda q ~\frac{\K \big(2 b (-2+m)+m V\big)}{6 b \E}
\\
 0 & 0 & 0 & 0 & 0
 \\
 0 & 0 & 0 & V + \frac{2 b \big(3 \E -\K (2 - m)\big)}{\K m} & 6
\end{pmatrix}
\end{align*}
with $V_1,V_2,V_3$ given by~\eqref{e:Vdef} as before and
\begin{align*}
&\gamma_1 = \lambda q~\frac{4 b^2 \big(3 \E (1+m)-\K(4-7m+m^2)\big)+2 b m (-3 \E+\K-2 \K m) V-\K m^2 V^2}{36 b (\E-\K) m}\,, \\
&\gamma_2 = \lambda q~\frac{4 b^2 \big(3 \E(1-2 m)+\K (-4+m+2 m^2)\big)+2 b m (-3 \E+\K+\K m) V-\K m^2 V^2}{36 b \big(\E-\K(1-m)\big) m}\,, \\
&\gamma_3 = \lambda q~\frac{4 b^2 \big(\K(8-8 m - m^2)-3 \E (2-m)\big)-2 b \big(3\E+2 \K (-2+m)\big) m V-\K m^2 V^2}{36 b \E m}\,.
\end{align*}
}
\unskip
Finally, to remove singularities from the system~\eqref{e:matrixeqn} one must first subtract the product of the {compatibility equation}~\eqref{e:constr} times the diagonal matrix {
$\mathrm{diag}(c_1,c_2,c_3,c_4,c_5)$,
with
\begin{align*}
&c_1 = \frac{\lambda q \sqrt{\beta}}{12 b} \, \frac{-2 b \big(\K (1-m)^2+\E (1+m)\big)+\big(\E-\K (1-m)\big) m V}{\E-\K}\,,  \\
&c_2 = \frac{\lambda q \sqrt{\beta}}{12 b} \, \frac{-2 b \big(\E(1- 2 m) + 2 \K (-1+m)\big)+\E m V}{\E-\K (1-m)}\,, \\
&c_3 = \frac{\lambda q \sqrt{\beta}}{12 b} \, \frac{2 b \big(\E (2-m)+\K (-2+m+m^2))+ \big(\E - \K (1-  m)\big) m V}{\E} \\
& -\lambda q~\frac{ \K (1-m) \sqrt{\beta} \big(2 b (1-2 m) + m V\big)}{12 b \big(\E-\K (1-m)\big)} \,, \\
&c_4 =  \frac{V+\lambda q^2}{2 \sqrt{\beta}} - m \sqrt{\beta} \frac{\K(1-m)}{\E - \K(1-m)}\,, \quad
c_5 = - 3 \sqrt{\beta} \frac{\E - \K(1-m)}{\K}\,,
\end{align*}
and $\beta = b/m$.
Then one need to subtract the product of the constraint~\eqref{e:pcontr} times the diagonal matrix
$\mathrm{diag}(d_1,d_2,d_3,0,0)$,
with
\begin{align*}
&d_1 = -\lambda q~\frac{\K \big(2 b (1+m)+m V\big)}{36 b (\E-\K)}\,, \qquad
d_2 = -\lambda q~\frac{\K q \big(2 b (1-2m)+m V\big)}{36 b \big(\E-\K (1-m)\big)} \,,\\
&d_3 = -\lambda q~\frac{\K \big(2 b (-2+m)+m V\big)}{36 b \E}\,.
\end{align*}}
Doing so yields the singularity-free system~\eqref{e:kpwhitham}.


\end{document}